\documentclass[twocolumn,trackchanges,resetfootnote]{aastex7}
\usepackage{xcolor,soul,color}
\usepackage[inline,shortlabels]{enumitem}
\usepackage{amsmath}
\usepackage{bm}
\usepackage{threeparttable}
\usepackage{algorithm,algorithmic}
\usepackage[export]{adjustbox}

\newcommand{\MZ}[1]{\textbf{\textcolor{blue}{[M:#1]}}}
\newcommand{\YDX}[1]{\textbf{\textcolor{magenta}{[YD:#1]}}}

\newcommand{\revise}[1]{\textcolor{black}{#1}}
\newcommand{\rerevise}[1]{\textcolor{black}{#1}}
\graphicspath{{./}{figure/}}
\defcitealias{furen}{D26}

\begin{document}

\title{Synthesis imaging with a lunar orbit array: III. {\bf A}ugmented lagrangian {\bf M}ultiplier {\bf I}maging using {\bf G}radient descent {\bf O}ptimization (\texttt{AMIGO})}

\correspondingauthor{Yidong Xu}
\email{xuyd@naos.cas.cn}

\author[0000-0002-2744-0618]{Meng Zhou}
\affiliation{State Key Laboratory of Radio Astronomy and Technology, National Astronomical Observatories, CAS, 20A Datun Road, Chaoyang District, Beijing 100101, People’s Republic of China}
\email{zhoumeng@bao.ac.cn}

\author[0000-0001-8075-0909]{Furen Deng}
\affiliation{State Key Laboratory of Radio Astronomy and Technology, National Astronomical Observatories, CAS, 20A Datun Road, Chaoyang District, Beijing 100101, People’s Republic of China}
\affiliation{School of Astronomy and Space Science, University of Chinese Academy of Sciences, Beijing 100049, People’s Republic of China}
\email{xxxx@bao.ac.cn}

\author[0000-0003-3224-4125]{Yidong Xu}
\affiliation{State Key Laboratory of Radio Astronomy and Technology, National Astronomical Observatories, CAS, 20A Datun Road, Chaoyang District, Beijing 100101, People’s Republic of China}
\email{xuyd@naos.cas.cn}

\author[0000-0001-6475-8863]{Xuelei Chen}
\affiliation{State Key Laboratory of Radio Astronomy and Technology, National Astronomical Observatories, CAS, 20A Datun Road, Chaoyang District, Beijing 100101, People’s Republic of China}
\affiliation{School of Astronomy and Space Science, University of Chinese Academy of Sciences, Beijing 100049, People’s Republic of China}
\affiliation{Center of High Energy Physics, Peking University, Beijing 100871, People’s Republic of China}
\email{xuelei@cosmology.bao.ac.cn}

\begin{abstract}
Ground-based radio observations below 30 MHz are severely limited by ionospheric interference and radio frequency interference (RFI) from Earth. A lunar-orbiting radio interferometer mission, the Discovering the Sky at the Longest wavelength (DSL, also known by its Chinese name ``Hongmeng''), has been proposed to overcome these obstacles. However, for such a mission, there are new challenges, such as the nearly all-sky field of view and dynamic 3D baselines, which require a huge computational cost for interferometric image reconstruction. 
In this work, we present \texttt{AMIGO} ({\bf A}ugmented lagrangian {\bf M}ultiplier {\bf I}maging using {\bf G}radient descent {\bf O}ptimization), a novel imaging algorithm tailored to lunar-orbiting arrays like DSL, combining the Mini-Batch Gradient Descent (MBGD) method with the Augmented Lagrangian Multiplier (ALM) technique.  MBGD reduces the computational complexity and memory cost, enabling efficient handling of large datasets. ALM \revise{flexibly} incorporates physical priors like non-negative sky temperature and prior angular power spectrum into the imaging algorithm, with adjustable stopping criteria to quantitatively control prior strength.
We validate \texttt{AMIGO} using mock visibility data generated under realistic DSL orbit configurations. Reconstructed sky maps at various frequencies and spatial resolutions show that this approach provides a computationally feasible framework for \revise{all-sky imaging} with a lunar-orbiting array.
\end{abstract}


\section{Introduction} \label{sec:intro}
Ground-based radio observations at frequencies below 30 MHz are severely blocked by ionospheric refraction and absorption. Additionally, both natural and artificial radio frequency interferences (RFI) strongly pollute astronomical signals in these low-frequency bands. To avoid the ionospheric effects and RFIs from the Earth, numerous concepts for space missions deployed on the lunar far-side surface or in lunar orbits have been proposed in recent decades. The proposed lunar surface missions include the ROLSES-1 \citep{Hibbard2026}, LuSee \citep{2023AGUFM.P31B..02B,LuSEE-Night2025mapmaking}, Astronomical Lunar Observatory (ALO, \citealt{2024AAS...24326401K}), the Large-scale array for radio astronomy on the farside (LARAF, \citealt{2024arXiv240316409C}),  Lunar Crater Radio Telescope (LCRT, \citealt{9438165}), FARSIDE \citep{2021RSPTA.37990564B} and FarView \citep{2024AdSpR..74..528P}.
The proposed lunar orbit missions include  the single satellite DARE \citep{2017ApJ...844...33B}, DAPPER \citep{2019AAS...23421202B}, PRATUSH \citep{2023ExA....56..741S}, SEAMS \citep{2021AIPC.2335c0005B}, CosmoCube \citep{Artuc2024,Zhu2025},
and the array Discovering the Sky at the Longest wavelength (DSL, \citealt{2021RSPTA.37990566C,Chen2023}). The orbit missions are simpler to implement from an engineering perspective. However, orbital interferometric observations will introduce additional systematic uncertainties related to dynamic baseline determination and inter-satellite wireless communication \citep{paperII}.

The DSL mission is one such proposed lunar orbit mission, comprising a satellite array with one mother satellite and nine daughter satellites \citep{2021RSPTA.37990566C,Chen2023}. The mother satellite collects data from the daughter satellites, performs cross-correlation analyses, and transmits the processed data back to Earth. One daughter satellite is designed to measure the global 21 cm spectrum across the 30–120 MHz band, while the remaining eight conduct global spectrum measurements below 30 MHz and interferometric all-sky imaging. The satellites follow a linear formation in a shared circular lunar orbit with precessional motion, generating three-dimensional baselines ranging from 100 m to 100 km.

Radio interferometric imaging using such a lunar orbital array differs significantly from ground-based arrays or even previous Earth-orbiting arrays. For the DSL mission, novel imaging algorithms are required to address challenges including an all-sky field of view, non-coplanar dynamic baselines, mirror symmetry relative to the orbital plane, and position-dependent sky blockage by the Moon. \cite{2018AJ....156...43H} proposed a brute-force map-making method that solves linear mapping equations relating sky intensity to visibilities. \cite{2022MNRAS.510.3046S} applied this method to a realistic DSL array configuration, presenting preliminary sensitivity estimates that account for thermal noise. To enhance algorithm performance under practical issues, \cite{furen} (hereinafter \citetalias{furen}) incorporated a “breathing” strategy -- where satellites periodically move closer to and farther from one another, and explored the impacts of the aliasing effect from sub-pixel noise and errors introduced by regularization parameter choices.
Besides interferometric imaging, the anisotropic distribution of the low-frequency spectrum can also be reconstructed in extremely low resolutions by using observations of a single antenna on lunar orbit \citep{2025ApJ...994..226A,2025arXiv250713102L}. 
However, these approaches become computationally prohibitive in computing time and memory for all-sky imaging with a high resolution and dynamic 3D baselines. 
Note that any imaging algorithm designed for a lunar orbital array is more general in the sense of problem complexity. 

In this work, we present \texttt{AMIGO} ({\bf A}ugmented lagrangian {\bf M}ultiplier {\bf I}maging using {\bf G}radient descent {\bf O}ptimization), a novel imaging algorithm tailored to lunar orbital radio interferometer arrays like DSL, combining the Mini-Batch Gradient Descent (MBGD, \citealt{8264077}) method with the Augmented Lagrangian Multiplier (ALM, \citealt{hestenes1969multiplier,Powell1969method,doi:10.1137/0312021}) technique.
Specifically, MBGD significantly reduces both computational time and memory costs, while ALM enables \revise{flexible incorporation of physical constraints and} quantitative control over prior strength through adjustable stopping criteria. 

The paper is organized as follows. In Section~\ref{sec:method}, we formulate the interferometric imaging optimization problem of \texttt{AMIGO} specific to the DSL mission and introduce the implementation of ALM and MBGD for its solution. In Section~\ref{sec:sim}, we describe the generation of mock data used to test \texttt{AMIGO}. In Section~\ref{sec:results}, we present reconstructed maps at various frequencies and spatial resolutions. In Section~\ref{sec:discuss} we compare \texttt{AMIGO} with previous approaches. Finally, we draw our conclusion in Section~\ref{sec:conclusions}.

\section{Methods} \label{sec:method}
Generally, in radio interferometric imaging, one observes visibility functions $\mathbf{V}^{\rm obs}$ from pairs of interferometer array elements with baselines $\mathbf{b}$. The visibility functions can be regarded as linear combinations of sky temperatures $\mathbf{s^{\rm true}}$ on pixels within the field of view, i.e., $\mathbf{V}^{\rm obs} = \mathbf{B}\,\mathbf{s}^{\rm true}+\bm{\eta}$, where $\mathbf{B}$ is the synthesis beam of the interferometer array, and $\bm{\eta}$ is the thermal noise. Due to incomplete {\it{uvw}} coverage and thermal noise, this set of linear equations cannot be solved perfectly. Therefore, we only expect to find the best ``compromise'' 
of $\mathbf{s^{\rm true}}$ from which the resimulated visibility functions $\mathbf{V^{\rm{sim}}}$ match the input $\mathbf{V}^{\rm obs}$. From this perspective, imaging can be regarded as an optimization problem that minimizes the difference between $\mathbf{V^{\rm{sim}}}$ and $\mathbf{V}^{\rm obs}$.

\subsection{Optimization Problem Set}
\label{subsec:problemset}
We first set up the optimization problem with respect to the radio interferometric imaging. A common way to quantify the differences between $\mathbf{V^{\rm{sim}}}$ and $\mathbf{V}^{\rm obs}$ is to define a data fidelity term in the least-squares sense, i.e.,
\begin{eqnarray}
\label{eqn:Jdata}
    J^{\rm data} &=& \sum_g\frac{|V_g^{\rm sim}-V_g^{\rm obs}|^2}{2\sigma_g^2}\,,
\end{eqnarray}
where $g$ denotes the index of visibility data and $\sigma_g$ is the thermal noise level of visibility functions. 

Optimizing this quadratic function with the Least Squares Method is equivalent to Maximum Likelihood Estimation (MLE) with a Gaussian likelihood, which has been well studied in recent decades. However, one may get an overfitted and incorrect solution by minimizing Equation~(\ref{eqn:Jdata}) in the presence of noise and incomplete $uvw$ coverage \citepalias{furen}.
A common strategy against overfitting is to introduce some denoising or regularization terms, e.g., reweighted total variation \citep{2023RAA....23l5017Y}, and Tikhonov regularization \citepalias{furen}, to suppress excess power at small scales. In Bayesian analysis, adding these terms is equivalent to imposing some {\it{prior}} knowledge. These {\it{priors}} set additional constraints on parameters to be reconstructed. \revise{Note that the priors here are not limited to explicit prior distributions of parameters, as in the Maximum A Posteriori (MAP) framework.} 

\revise{In this work, we impose a constraint based on the all-sky angular power spectrum, $C_\ell$. By definition, for one sky map, $C_\ell$ is given by
}
\begin{eqnarray}
    C_\ell = \frac{1}{2\ell+1}\sum_{m=-\ell}^\ell\left|a_{\ell m}\right|^2\,,
\end{eqnarray}
\revise{where $\ell$ denotes the multipole index, $m$ denotes the azimuthal index, and $a_{\ell m}$ is spherical harmonic expansion coefficients of the sky map. Then the constraint is set by}
\begin{eqnarray}
    H_\ell = \frac{C_\ell^{\rm sim}}{C_\ell^{\rm prior}}-1=0\,,
\end{eqnarray}
where $C_\ell^{\rm sim}$ is the angular power spectrum of the \revise{resimulated} sky map\revise{, and $C_\ell^{\rm prior}$ is the prior one.} \revise{The main advantage of using it as a constraint, unlike total variation, is that the prior information from the angular power spectrum has a direct {\it physical interpretation}, i.e., it represents the power on different angular scales.} Furthermore, as statistics, a prior angular power spectrum is more reliable than a prior sky map extrapolated from high-frequency observations.

We also include an inequality constraint that the sky temperature should always be positive, i.e.,
\begin{eqnarray}
    G_n=s_n\geq 0\,,
\end{eqnarray}
where $n$ denotes the index of the $n$-th pixel in the sky. \revise{This non-negative sky temperature constraint is physically motivated by the intrinsic brightness of celestial sources, although it may introduce mild positive biases in low-signal regions.}

Therefore, we have set up a constrained optimization problem:
\begin{eqnarray}
    {\rm{min}}_s\left\{J^{\rm data}\right\}\ {\rm s.t.}\ H_\ell=0\,,G_n\geq0\,.
\end{eqnarray}

\subsection{Augmented Lagrangian Multiplier}
It is \revise{generally} more difficult to solve a constrained optimization problem than an unconstrained one. \revise{A common strategy is to} use the Lagrangian Multiplier or its variants to convert the constrained optimization problem into an unconstrained one. One of its variants is the Augmented Lagrangian Multiplier (ALM). ALM improves the traditional Lagrangian Multiplier Method by incorporating penalty terms, thereby strengthening the handling of constraints in optimization problems. It combines the constraint-characterizing capability of Lagrange multipliers with the numerical stability of penalty function methods, reducing the sensitivity to step sizes during iterations. ALM and other variants of the Lagrangian Multiplier Method find extensive applications in machine learning, engineering optimization, and other fields where constrained optimization is required, including imaging \citep{2010ITIP...19.2345A,2011ITIP...20..681A}. Therefore, for the constrained problem in Section~\ref{subsec:problemset}, we use ALM to convert it into the following unconstrained optimization problem,

\begin{widetext}
\begin{eqnarray}
\label{eqn:model}
    {\rm min}_s\left\{J^{\rm tot} = J^{\rm data}+\frac{\rho_1}{2}\sum^{\ell_{\rm max}}_{\ell=\ell_{\rm min}}( H_\ell+\frac{\lambda_\ell}{\rho_1})^2+\frac{\rho_2}{2}\sum_n [{\rm min}(G_n+\frac{\mu_n}{\rho_2},0)]^2\,\right\}.
\end{eqnarray}
where we denote the overall cost function as $J^{\rm tot}$. Here $\lambda_\ell$ and $\mu_n$ 
are multipliers for $H_\ell$ and $G_n$, respectively. $\rho_1$ and $\rho_2$ are penalty factors that can control the strength of constraints. When they approach $+\infty$, one can obtain a solution that perfectly matches constraints. If they are fixed to zero, then the solution should only be determined by the data fidelity term. 
\revise{In principle, the general framework is flexible enough to accommodate additional constraints. We will also combine other constraints and assess their improvements to imaging quality in future work.}

\revise{Following the standard ALM solution procedure}, we can decompose it into three sub-iterations. \revise{The update rule for the $k$-th iteration} with respect to Equation~(\ref{eqn:model}) can be given by,
\begin{eqnarray}
    \label{eqn:updating}
    {\bf{s}}^{k+1}= {\rm argmin_{\bf{s}}} [(J^{\rm tot})^k]\nonumber\\
    \lambda_\ell^{k+1} = \lambda^{k}_\ell+\rho_1 H_\ell^{k+1}\nonumber\\
    \mu^{k+1}_n = {\rm min}(\rho_2 G^{k+1}_n+\mu^k_n,0)
\end{eqnarray}
\revise{Here, $H_\ell^{k+1}$ is updated using ${\bf{s}}^{k+1}$.}

\end{widetext}
The first sub-iteration is a sub-optimization problem. It can be easily solved because $J^{\rm tot}$ is differentiable. Since the entire system is quadratic to sky temperature at most, we do not need too many iterations to meet the global stopping criteria. 

For fixed values of $\rho_1$ and $\rho_2$, when the global stopping criteria are met, the constraint equations may still have some residuals, which can be defined as,
\begin{eqnarray}
    \Delta_H &=& \left(\sum_{\ell=\ell_{\rm min}}^{\ell_{\rm max}} H_\ell^2/N_\ell\right)^{1/2}\,;\nonumber\\
    \Delta_G &=& {\rm max}_n[|{\rm min}(G_n,0)|/\sigma_s]\,,
\end{eqnarray}
where $N_\ell$ is the number of \revise{multipoles} from $\ell_{\rm min}$ to $\ell_{\rm max}$, 
and $\sigma_s$ is the standard deviation of the prior map. Here, we use the full range of $\ell$ modes given the target spatial resolution of the reconstructed map. In principle, one can also use part of them to optimize the reconstruction quality on certain scales.
$\Delta_H$ and $\Delta_G$ quantify the ``closeness'' between the reconstructed maps and the priors in the sense of \revise{constraint satisfaction}. \revise{Increasing $\rho_1$ and $\rho_2$ effectively strengthens the priors and, yielding smaller} $\Delta_H$ and $\Delta_G$, respectively. Therefore, the strength of these priors is {\it controllable} by setting stopping criteria based on thresholds of $\Delta_H$ and $\Delta_G$. \revise{In practice, we start with some initial values of $\rho_1$ and $\rho_2$ and progressively increase them until the residuals fall within the thresholds}\footnote{To save computational costs, \revise{we only increase $\rho_1$ and $\rho_2$ when the residuals do not decrease anymore}. Therefore, the residuals  
can be lower than the expected thresholds. \revise{ We also require that the final values of $\rho_1$ and $\rho_2$ must be at least 100 times the initial ones to prevent excessively large initial values. Otherwise, we will restart with some smaller values.}}. We vary the threshold values and explore their impact on imaging quality in Section~\ref{subsec:priortest}.

\subsection{Mini-Batch Gradient Descent}
The first sub-iteration \revise{of the ALM scheme}, i.e., Equation~(\ref{eqn:updating}), is similar to the optimization problem in \cite{2018AJ....156...43H},  \cite{2022MNRAS.510.3046S}, and \citetalias{furen}, \revise{as well as} the first sub-problem in \cite{2023RAA....23l5017Y}. In principle, it can be solved \revise{via matrix inversion} using the split Bregman method \citep{doi:10.1137/080725891,doi:10.1137/090753504} or other variants of the Alternating Direction Method of Multipliers (ADMM, \citealt{eckstein1992douglas}). However, the biggest limitation for matrix inversion is its huge computational cost. Moreover, for the DSL mission \citep{2022MNRAS.510.3046S}, one must load a huge \revise{response} matrix $\mathbf{B}$, due to its large number of pixels for the whole sky, and huge amount of visibility data generated during a full precession cycle. 

A feasible solution to reduce the computational cost is to use the mini-batch gradient descent method. The general idea of any gradient-descent-like method is to iteratively determine a descent direction from the gradients of the cost function and to perform a one-dimensional line search along this direction in the parameter space. Therefore, the memory cost is reduced from $O(N^2)$ to $O(N)$, where $N$ is the parameter space size, in this case, the number of sky map pixels. \revise{The time cost per iteration is $O(MN)$, where $M$ is the sample size (the number of visibilities).} The number of iterations required to reach convergence is typically less than $O(N)$ for a linear system such as radio interferometric imaging. Therefore, the overall time cost should be well below $O(MN^2)$, and significantly less than $O(N^3+MN^2)$ as required by matrix inversion with the Least Squares Estimation. Specifically, Mini-Batch Gradient Descent (MBGD) bridges the gap between two extreme cases: full-batch Gradient Descent (BGD) and Stochastic Gradient Descent (SGD). Traditional BGD uses the full dataset for stable but computationally heavy updates, while SGD relies on individual data points, being efficient but noisy. MBGD balances this tradeoff by using small data subsets. It can reduce the variance of updates for steadier convergence and retain efficiency, which is ideal for large datasets.

In this work, we split the full set of visibility data into mini-batches of size $N_{\rm mini}$. In {\it one} epoch of the first sub-iteration of Equation~(\ref{eqn:updating}), 
for {\it each} mini-batch, we accumulate the value of the cost function $J^{\rm tot}$, \revise{evaluate the gradients $\partial J^{\rm tot}/\partial s_n$ with respect to the sky temperature (see Appendix~\ref{app:gradients}  for detailed derivations),} and update $s_n$ via
\begin{equation}
    s_n\leftarrow s_n-\frac{\alpha}{M}\frac{\partial J^{\rm tot}}{\partial s_n}
\end{equation}
where $\alpha$ is the learning rate. \revise{We adopt a simple backtracking-like \citep{nocedal2006numerical} learning-rate update rule. A similar scheduler technique, {\tt ReduceLROnPlateau}, which monitors the loss function and adjusts the learning rate accordingly, has been implemented in PyTorch \citep{paszke2019pytorch} and TensorFlow \citep{tensorflow2015whitepaper}, and is widely used in machine learning. We set the initial value of $\alpha$ to be 1.0. The learning rate is then decayed by a factor of 0.1, and the updated $s_n$ is rejected if the accumulated $J^{\rm tot}$ does not decrease in some epoch, until $\alpha$ falls below $10^{-7}$. To accelerate convergence, we also increase $\alpha$ by a factor of 1.2 if the accumulated $J^{\rm tot}$ does not increase in some epoch.}

\revise{In addition to shuffling mini-batches, as is common practice in machine learning}, we also try fixed orderings, e.g., an ascending or descending order of baseline length. We examine the effect of various batch configurations on imaging quality in Section~\ref{subsec:batchtest}.

\subsection{Aliasing Effect}
In imaging and signal processing, aliasing is a critical distortion phenomenon that arises from an insufficient sampling rate with respect to the Nyquist criterion. In \revise{traditional 2D} interferometric imaging, the aliasing effect manifests as artifacts such as spurious periodic structures, blurred boundaries, or misrepresented fine details when using baselines beyond the Nyquist limit $b_{\rm NQ}=0.5\lambda/\theta_p$, where $\theta_p$ is the target pixel size. \revise{For the DSL mission, longer 3D baselines introduce aliasing in directions where projected baselines are longer than $b_{\rm NQ}$, but can improve imaging quality along directions where projected components remain within $b_{\rm NQ}$. To incorporate the benefits of longer baselines but still suppress aliasing effect, one possible solution is to adopt a pixel-averaging strategy on the response matrix ${\bf B}$ \citepalias{furen}.}

\revise{In principle, the same strategy could be adopted in {\tt AMIGO} to incorporate improvements from long baselines. However, to take full advantage of its lower computational cost, a better way is to reconstruct maps at a higher resolution determined by the Nyquist criterion. In this case, information sampled by longer baselines can be fully utilized without being smoothed out by the pixel averaging. If necessary, the reconstructed maps can also be downgraded to a lower resolution for a fair comparison with \citetalias{furen} and for validating improvements from longer baselines (see Section~\ref{subsec:comparison} for detailed discussion).}

\section{Simulation}
\label{sec:sim}
We use mock visibility data to verify \texttt{AMIGO}. In this section, we describe our approach to simulating the visibility functions.
\subsection{Simulating the visibility}
We adopt the same modeling of visibility functions for a lunar-orbited array in previous work \citep{2018AJ....156...43H,2022MNRAS.510.3046S,furen}. The visibility $V_{ij}$ measured by the elements $i$ and $j$ at a frequency of $\nu$ can be written as
\begin{equation}
    V_{ij}(t) = \int s(\hat{\mathbf{n}})S_{ij}(\hat{\mathbf{n}})A_{ij}(\hat{\mathbf{n}}){\rm exp}\left(-2\pi {\mathrm{i} }\hat{\mathbf{n}}\cdot\mathbf{r}_{ij}\,\nu/c\right)d\hat{\mathbf{n}}^2\,,
    \label{eqn:vis}
\end{equation}
where $\hat{\mathbf{n}}$ is the sky direction, $\mathbf{r}_{ij}$ is the baseline vector, $s(\hat{\mathbf{n}})$ is the sky signal, $S_{ij}(\hat{\mathbf{n}})=S_{i}(\hat{\mathbf{n}})S_{j}(\hat{\mathbf{n}})$ is the shading function, $A_{ij}(\hat{\mathbf{n}})=\sqrt{A_{i}(\hat{\mathbf{n}})A_{j}(\hat{\mathbf{n}})}$ is the power response of the primary beam.

We model the Moon as a fully opaque sphere and neglect its radiation, diffraction, and reflection. The shading function $S_{i}$ masks out the contributions from the sky directions blocked by the Moon, i.e. 
\begin{equation}
    S_{i}(\hat{\mathbf{n}}) = 
    \begin{cases}
    0,  &-\hat{\mathbf{n}}\cdot \mathbf{r}_{i}>\sqrt{|\mathbf{r}_{i}|^2-R_{\rm moon}^2} \\
    1, &{\rm else}
    \end{cases},
\end{equation}
where $R_{\rm moon}=1737.1\,{\rm{km}}$ is the radius of the Moon.

The DSL mission proposes to use three orthogonal pairs of short dipole antennas \citep{2021RSPTA.37990566C}, whose primary beam can be approximated as \citep{2022MNRAS.510.3046S}:
\begin{equation}
    A_i(\hat{\mathbf{n}})=1-(\hat{\mathbf{n}}\cdot\hat{\mathbf{n}}_i^a)^2\,,
\end{equation}
where $\hat{\mathbf{n}}_i^a$ is the antenna direction. For a demonstration of the algorithm, in this work we consider only one pair of antenna, and assume it is pointed to the Center of the Moon.

The thermal noise $\eta_{ij}$ on $V_{ij}$ can be modeled as a white noise. Its real part and imaginary part should be independent and follow a Gaussian distribution respectively, with a standard deviation given by
\begin{equation}
    \sigma_{ij}=\frac{T_{ij}^{\rm sys}}{\sqrt{2\Delta\nu t_{\rm int}}}\,,
\end{equation}
where $T_{ij}^{\rm sys} = \sqrt{(T^{\rm rec}+T_i^{\rm sky})(T^{\rm rec}+T_j^{\rm sky})}\approx (T^{\rm rec}+T_i^{\rm sky})$. Here we assume the receiver temperature $T^{\rm rec}$ is identical for each antenna. $T_i^{\rm sky}$ is the global sky temperature observed by the antenna $i$. The bandwidth is set at $\Delta\nu=8\,{\rm kHz}$. The actual data will be collected at fixed time intervals, but to reduce the amount of computation, we can re-bin the data to a resolution needed for accurate computation. We use the same criteria in \revise{\citetalias{furen}} to dynamically determine the {\it rebinned} integration time $t_{\rm int}$ for the post-processing on the ground, i.e.,
\begin{eqnarray}
    t_{\rm int} = {\rm min}(\frac{\lambda}{8|v_{ij}|},\,25s)\,.
\end{eqnarray}
where $\lambda$ is the wavelength and $|v_{ij}|$ is the changing rate of baseline length.

\subsection{Orbit Configurations}
We follow the same assumption in \citet{2018AJ....156...43H}, \citet{2022MNRAS.510.3046S}, and \citetalias{furen} that all satellites move on the same orbit without deviation, with a constant speed for most time and instantaneous velocity changes when necessary. The orbit of the whole satellite array is modeled as a combination of a uniform and stable circular motion at a height of 300\,km and an inclination angle of 30$^\circ$, and a uniform precession of the orbital plane with a period of 1.3 years. We only consider the observation time when the Moon blocks the Earth so that we can avoid the RFI from the Earth. We also adopt a {\it breathing} strategy in which the eight daughter satellites move closer and further away from each other periodically covering baselines from 100 m to 100 km, to obtain a better {\it uvw} coverage and correct for natural velocity deviation due to the inhomogeneous gravitational potential in lunar orbit.

\begin{table}[!h]
    \centering
    \caption{The positions of the satellites with respect to the first one at the beginning of each phase.}
    \begin{threeparttable}
    \begin{tabular}{c|c}
    \hline
      Time (day) & $r_n$ \\
      \hline
        0 &  $r^0_n$ \\
        7 &  $r^0_n/\mathcal{R}$ \\
        14 &  $r^0_n$ \\
        \hline
    \end{tabular}
    \begin{tablenotes}
    \footnotesize
    \item \textbf{Note.} Here $r_n^0$ is the initial relative position of the satellite $n$ with respect to the satellite 1.
    \end{tablenotes}
    \end{threeparttable}
    \label{tab:breathing}
\end{table}

In Table~\ref{tab:breathing}, we list the relative positions of the satellites with respect to the first one at the beginning of each phase for a better illustration.
We adopt the same parameterization of $r_n^0$ as in \citetalias{furen}, i.e.
\begin{equation}
\label{eqn:initial_position}
    r_n^0=b_0+a_0q_0^{n-2}\ \ \ \ \  {\rm for}\ \ n\geq2\,.
\end{equation}

We set $r_1^0=0\,$m and $r_8^0=100\,$km. To avoid the risk of collision, we also set the shortest baseline during one breathing period to 100\,m and the second shortest one to 500\,m. Therefore, given the compression ratio $\mathcal{R}$, one can get the parameters $(a_0,\ b_0,\ q_0)$ by plugging the above requirements into Equation~(\ref{eqn:initial_position}). In this work, we adopt an array configuration with $\mathcal{R}=10$. Note that the orbit configuration might affect the quality of imaging. This breathing parameterization may not be optimal. However, we have verified that, within \revise{safety constraints}, baseline distributions of various orbit configurations do not change dramatically. Additionally, the impact of orbit configurations should not be coupled with the systematics of the imaging algorithm. Searching for the optimal orbit configuration is not within the scope of the present paper.

\section{Results}
\label{sec:results}
In this section, we present the reconstructed maps obtained with our fiducial strategy of prior control and batch configuration. We use the angular power spectrum of the input map as the prior and set the fiducial prior threshold to $\Delta_H^{\rm thres}=0.01$ and $\Delta_G^{\rm thres}=0.01$ at $f=3.0$ and $10.0$ MHz. Since the $uvw$ coverage at $30.0$ MHz is worse, we adopt a looser threshold for better convergence, setting $\Delta_H^{\rm thres}=0.03$ and $\Delta_G^{\rm thres}=0.01$. In the fiducial batch configuration, we set the batch size to $N_{\rm mini}=2^{18}$ and arrange the visibility data batches in an ascending order (i.e., from short to long baselines). 

\begin{figure*}[!h]
\centering
\includegraphics[width=\columnwidth]{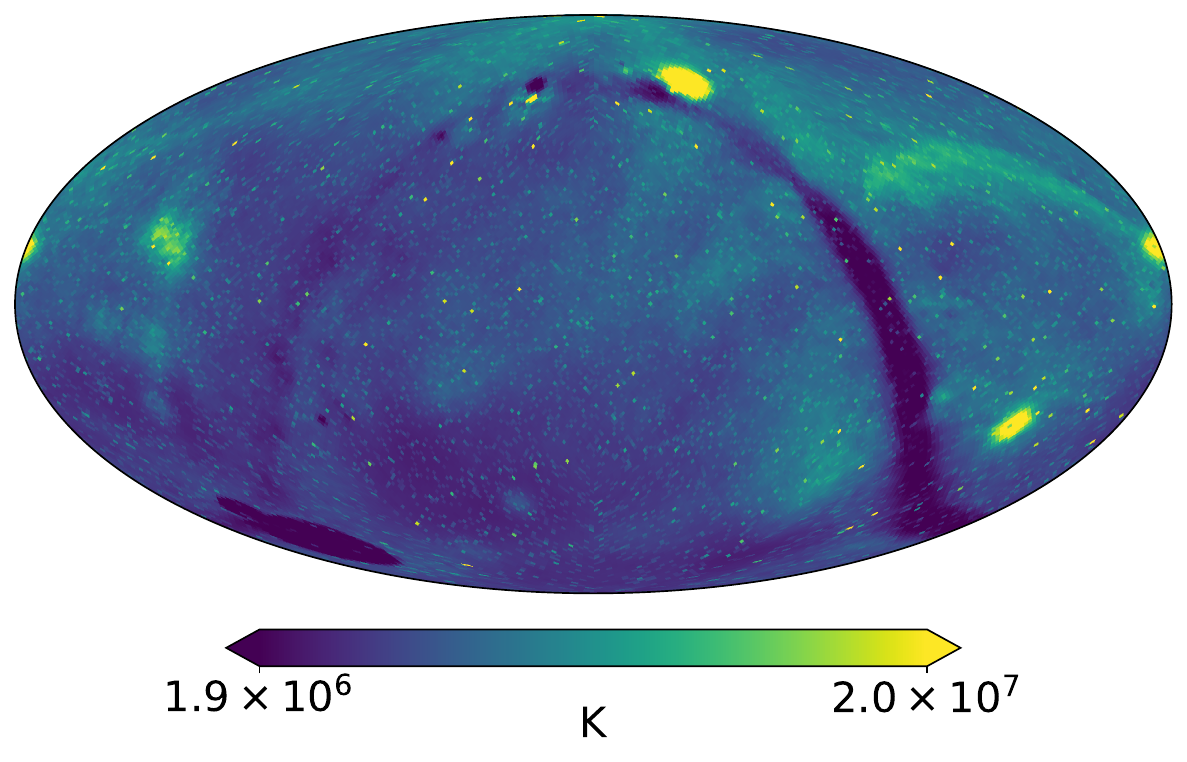}\\
\includegraphics[width=\columnwidth]{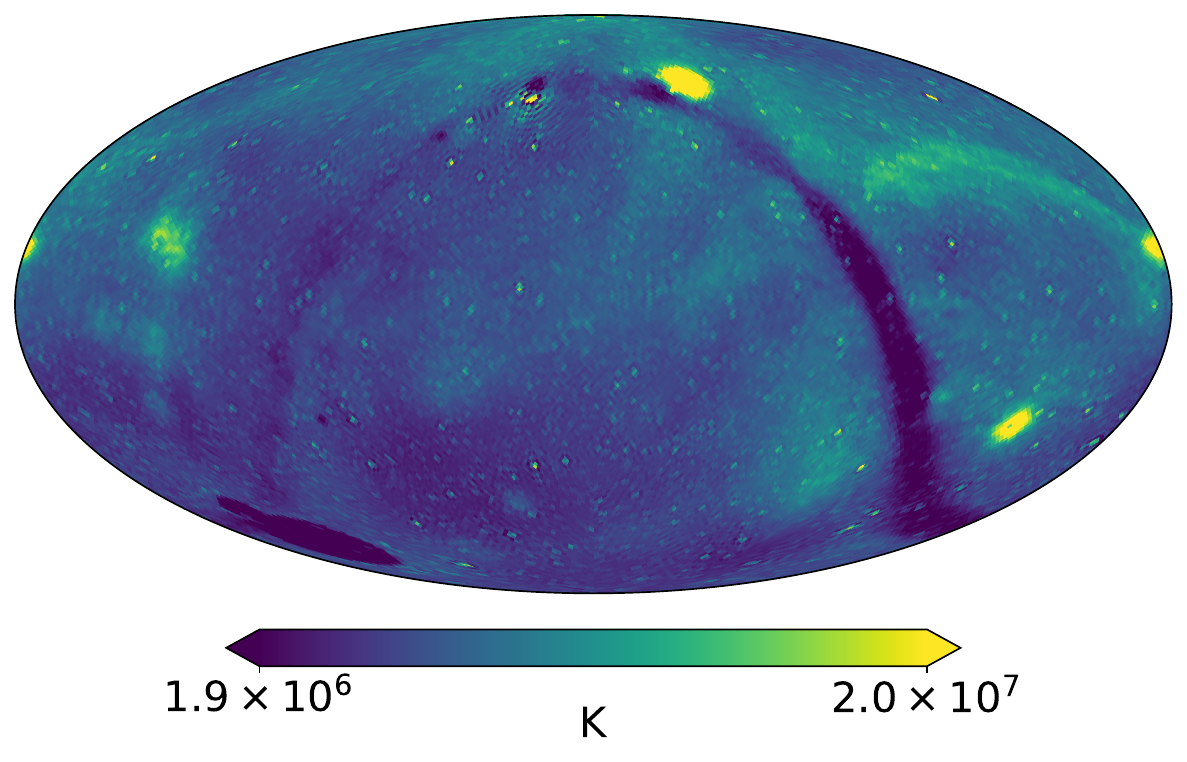}
\includegraphics[width=\columnwidth]{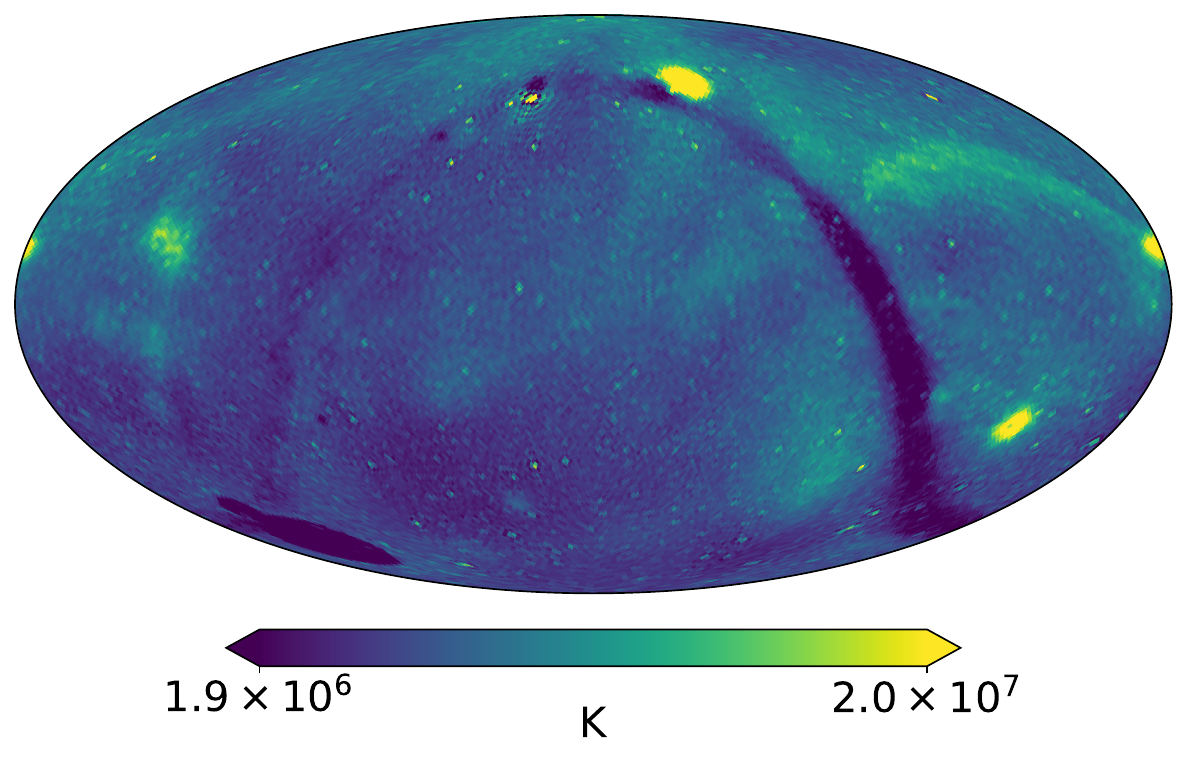}\\
\includegraphics[width=\columnwidth]{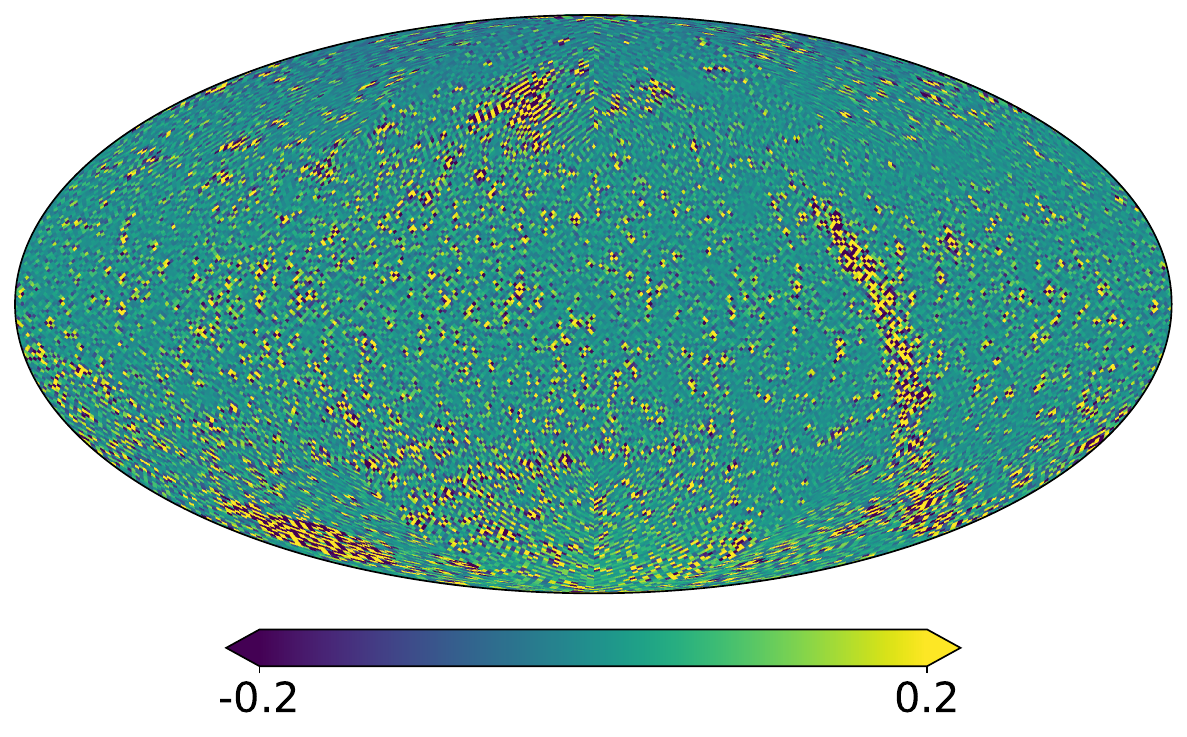}
\includegraphics[width=\columnwidth]{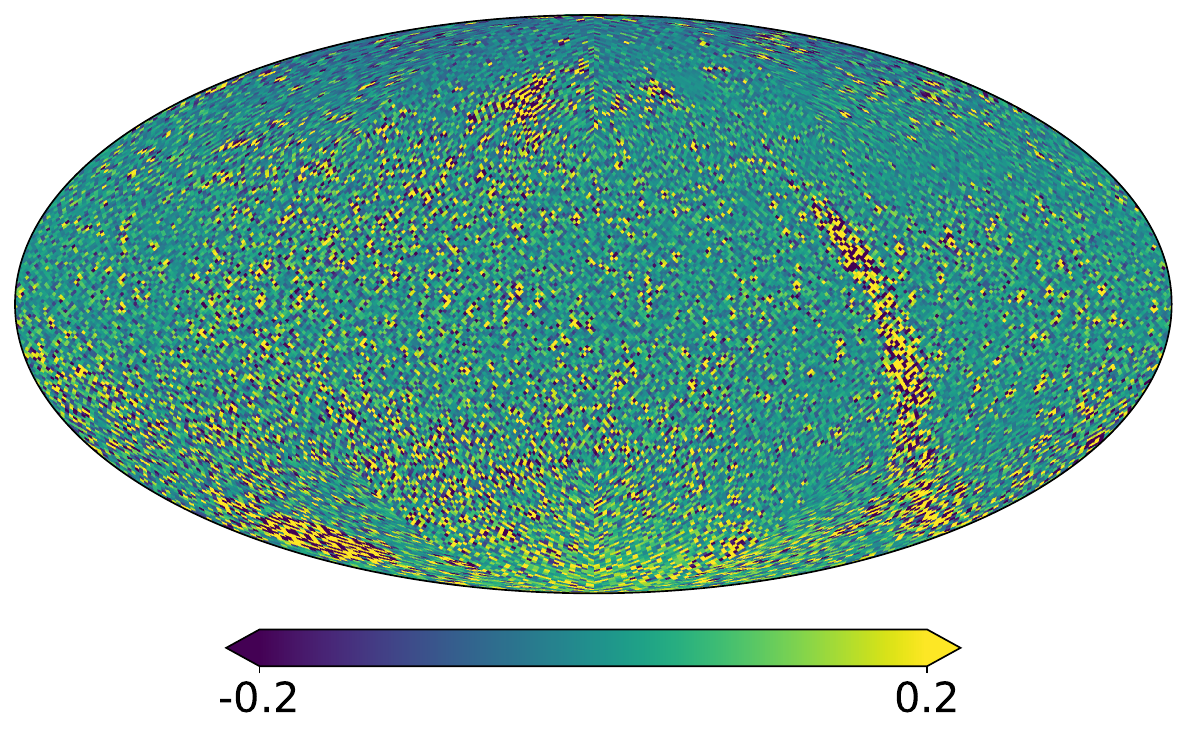}
\caption{Reconstructed maps at 3.0 MHz with NSIDE = 64. We plot the reconstructed sky maps without (Left) and with (Right) thermal noise
in the middle panels, and the fractional residuals in the bottom panels. We also plot the input map downgraded to NSIDE = 64 in the top panel for comparison.}
\label{fig:resolution_test_NSIDE64}
\end{figure*}

For NSIDE = 64, we start from a flat initial guess\revise{, $s^{\rm ini}_n=s^0$,} to solve the optimization problem in Equation~(\ref{eqn:model}). \revise{In the fiducial configuration, $s^0$ is set to an unbiased value, i.e., the mean value $\bar{s}^{\rm true}$ of the input sky map. We also discuss the convergence behavior over initial guesses in Section~\ref{subsec:convergence}.} We first reconstruct skymaps at 3.0 MHz with NSIDE = 64. We plot the reconstructed skymaps in Figure~\ref{fig:resolution_test_NSIDE64} in Ecliptic coordinates without (Left) and with (Right) thermal noise in the middle panels, and the fractional residuals in the bottom. We also plot the input map downgraded to NSIDE=64 in the top panel for comparison. We find that most fractional errors for the map are within 20\%, which is consistent with \citetalias{furen}. \revise{We observe the expected positive bias confined to low-signal regions. Since such regions should only account for a small fraction of the sky map, the overall impact remains negligible.}

\begin{figure*}[ht]
\centering
\includegraphics[width=\columnwidth]{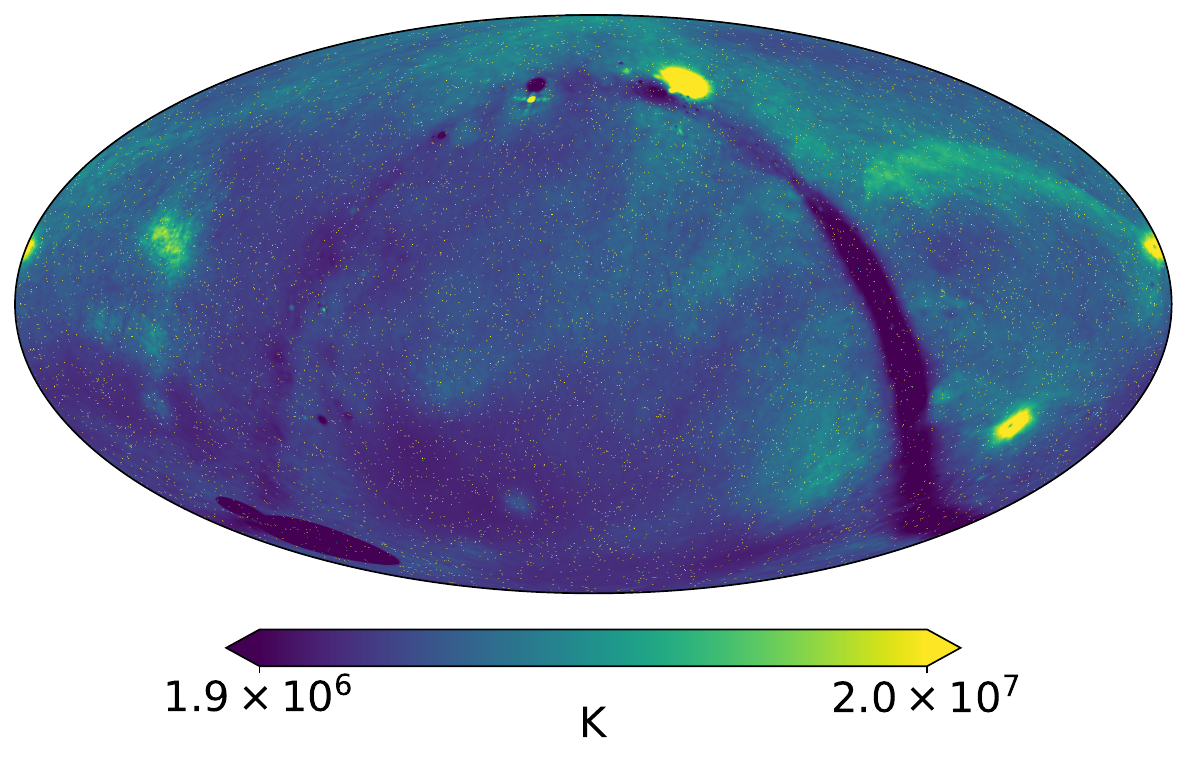}\\
\includegraphics[width=\columnwidth]{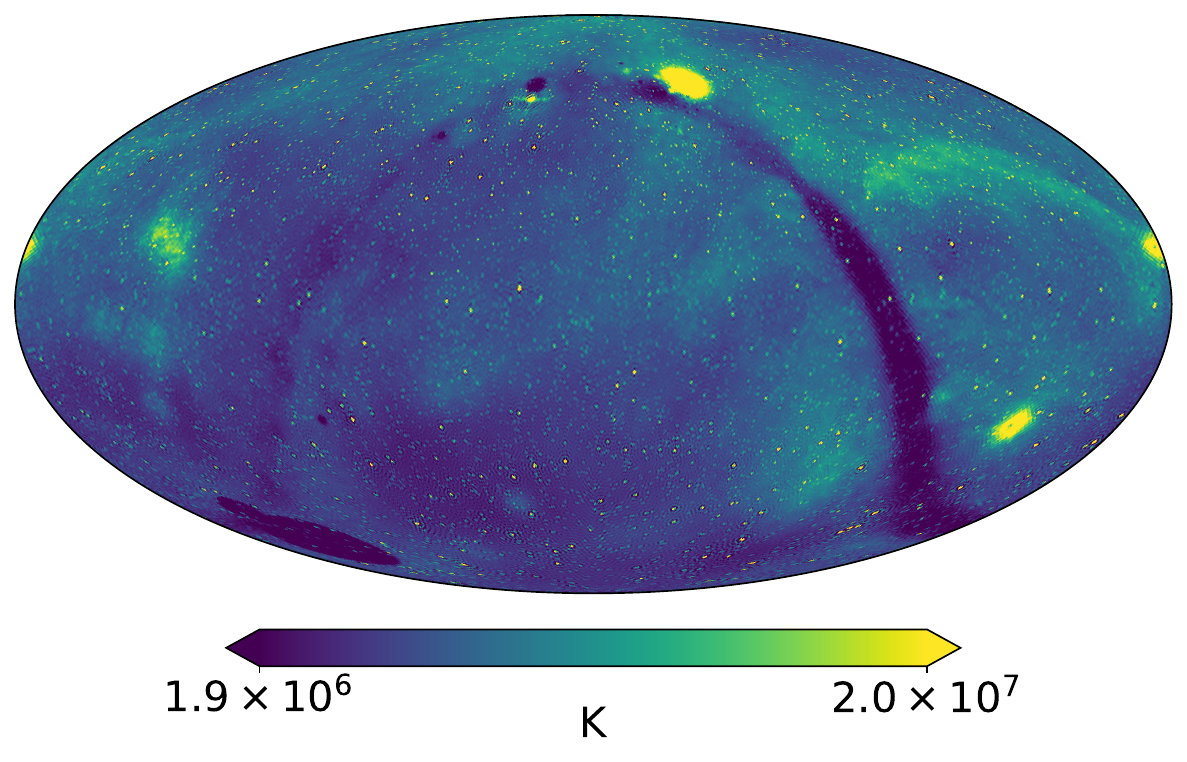}
\includegraphics[width=\columnwidth]{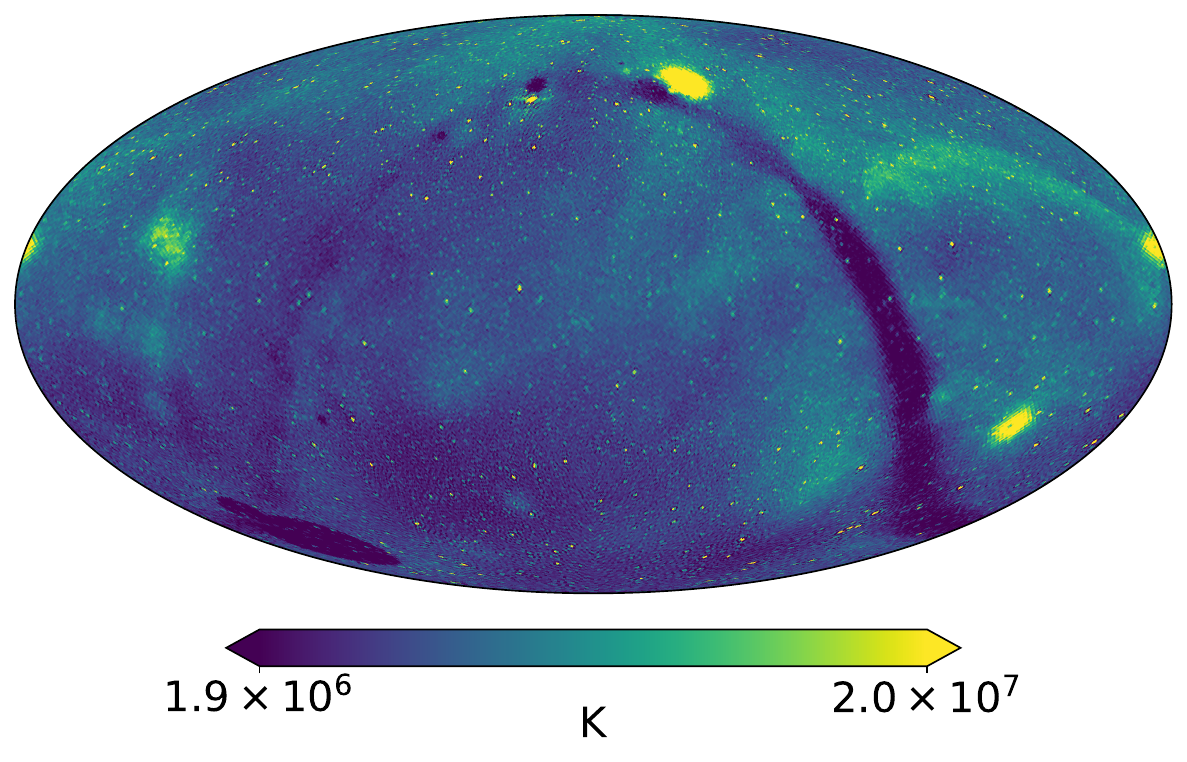}\\
\includegraphics[width=\columnwidth]{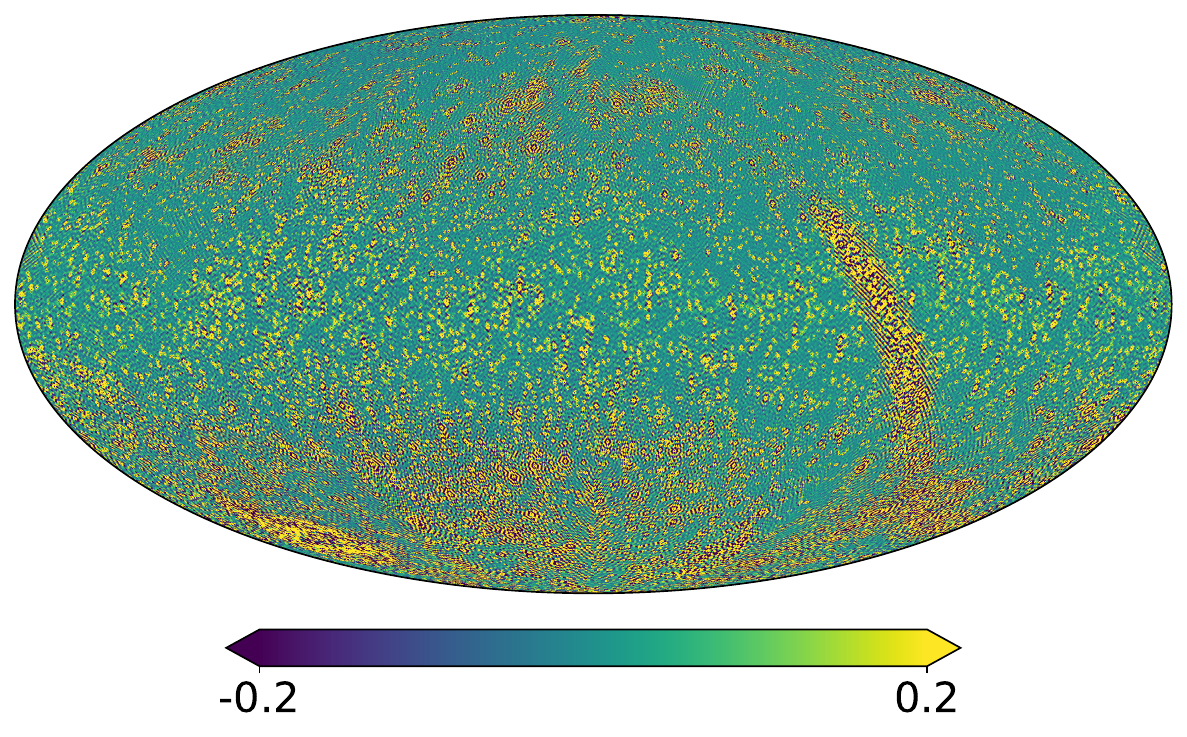}
\includegraphics[width=\columnwidth]{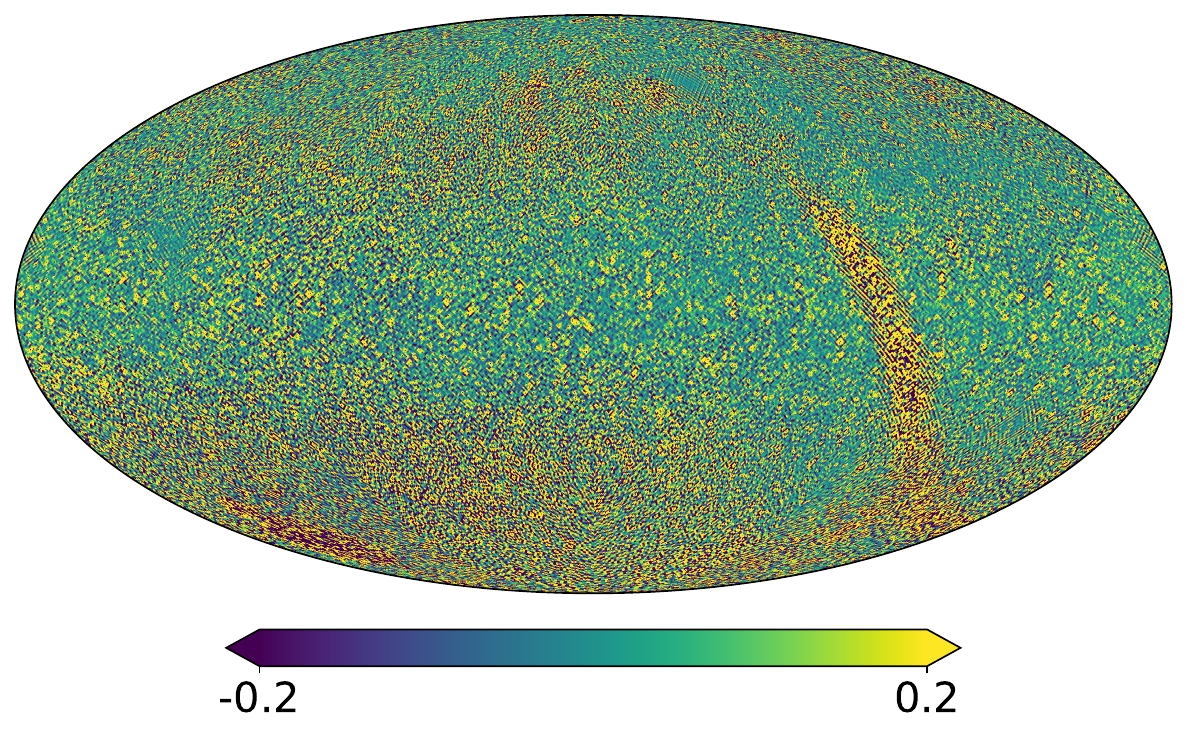}
\caption{Same as Figure~\ref{fig:resolution_test_NSIDE64} but with NSIDE=256.}
\label{fig:resolution_test_NSIDE256}
\end{figure*}

For higher resolutions, we take the reconstructed skymaps at lower resolutions as initial guesses in a hierarchical way. \revise{For example, we upgrade the reconstructed skymaps with NSIDE = 64 in Figure~\ref{fig:resolution_test_NSIDE64} to NSIDE = 128 as the initial guesses for maps with NSIDE = 128. After obtaining reconstructed maps with NSIDE=128, we upgrade them to serve as initial guesses for NSIDE = 256.} \revise{The hierarchical initial guess substantially reduces the computational cost, as a low-resolution sky map is much closer to the result with the global minimum of the cost function than a flat one. Meanwhile, it circumvents part of the local minima, thereby improving the overall convergence behavior (see Section~\ref{subsec:convergence}).} We plot the reconstructed skymaps at 3.0 MHz with NSIDE = 256 in Figure~\ref{fig:resolution_test_NSIDE256} in a similar way to Figure~\ref{fig:resolution_test_NSIDE64}. 
We find that although diffuse structures and part of point sources have also been well reconstructed, the fractional residuals increase significantly at higher resolution. This is consistent with our expectation because the confusion noise scales with $\theta_p^{-2}$. A few bright point sources on the reconstructed map appear to be more prominent than those on the input map. This is a visual \revise{artifact} in the reconstructed map, where the point source spreads to several adjacent pixels. Those brightest sources with temperatures exceeding the displayed upper limit of the colorbar can look larger than their actual sizes in the input map.

\begin{figure*}[!h]
\centering
\includegraphics[width=\columnwidth]{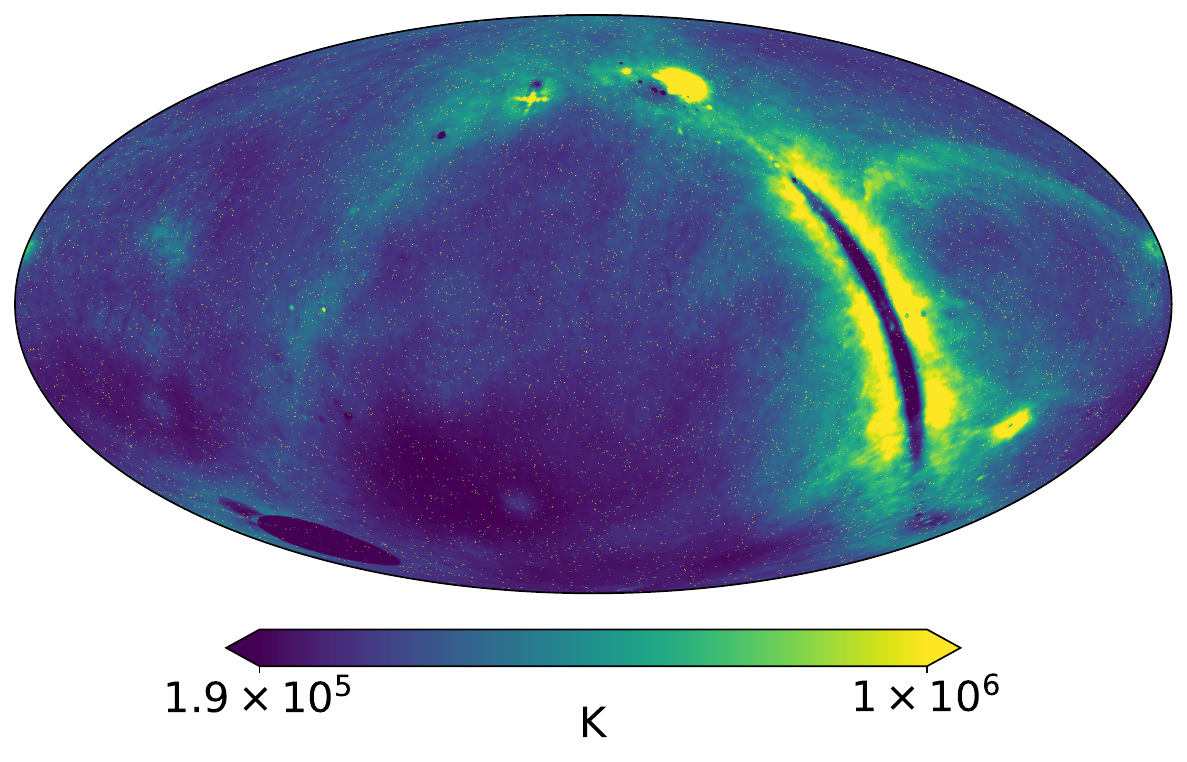}\\
\includegraphics[width=\columnwidth]{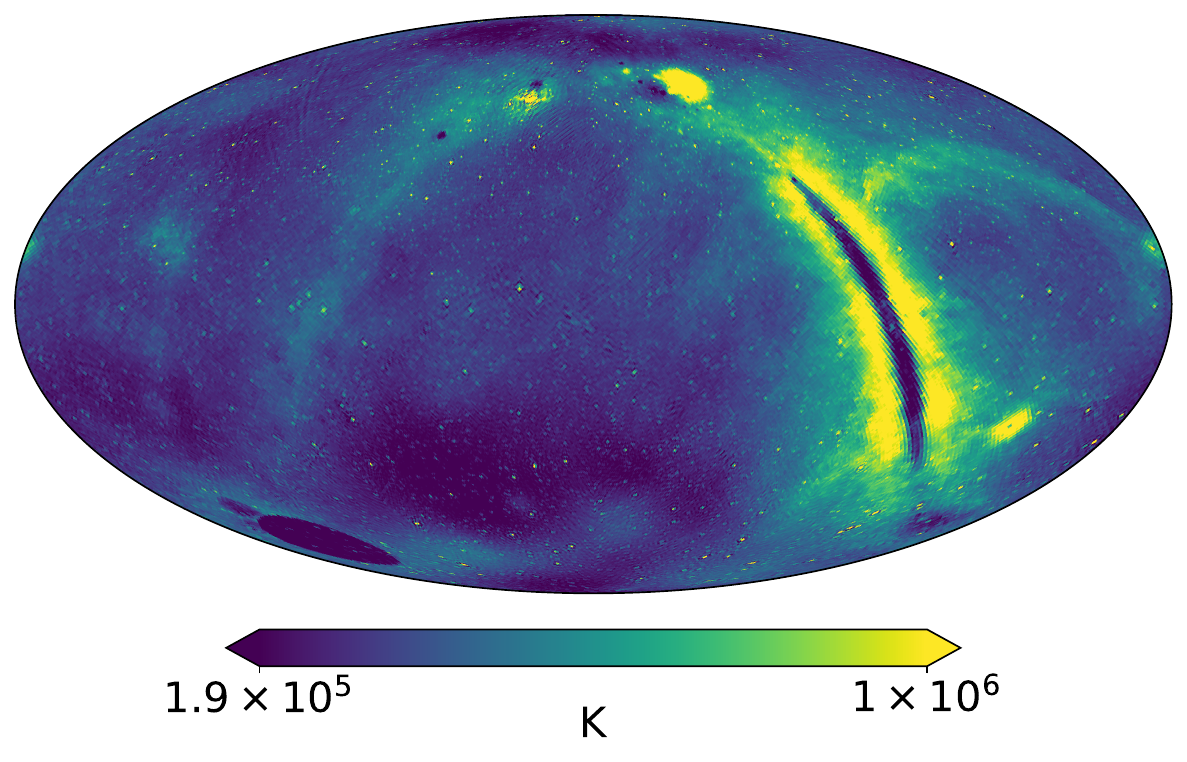}
\includegraphics[width=\columnwidth]{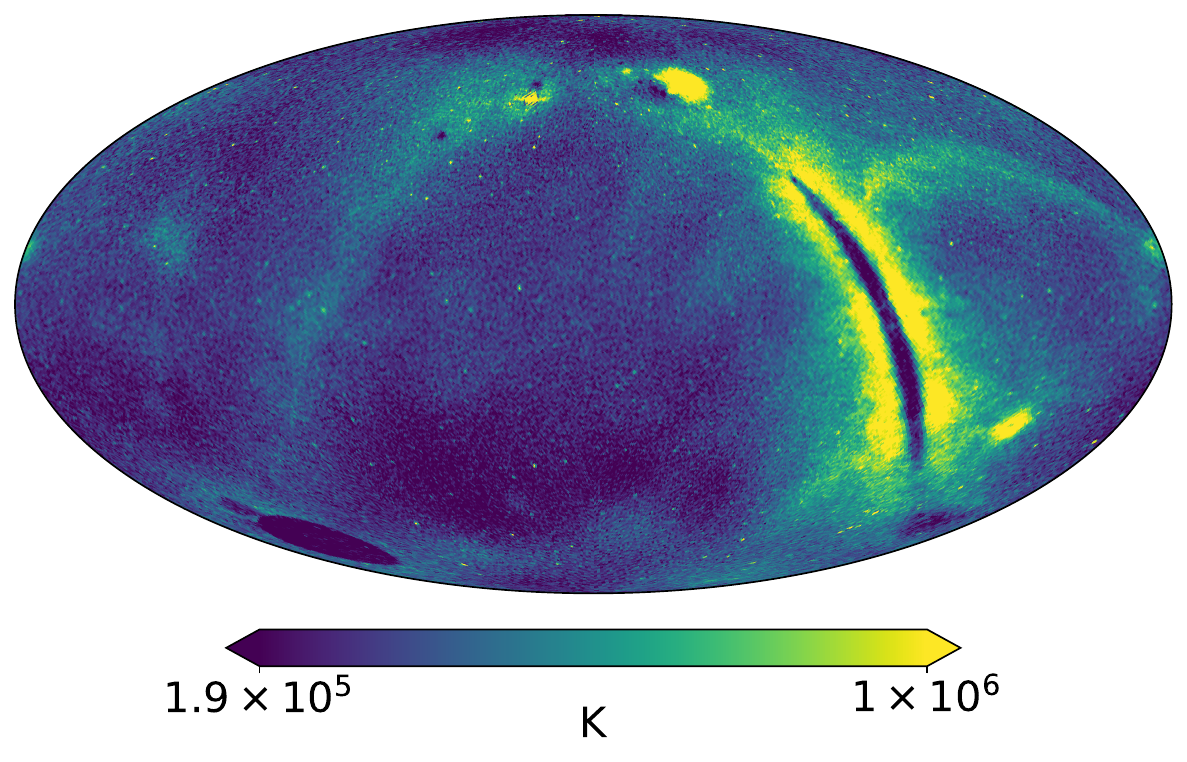}\\
\includegraphics[width=\columnwidth]{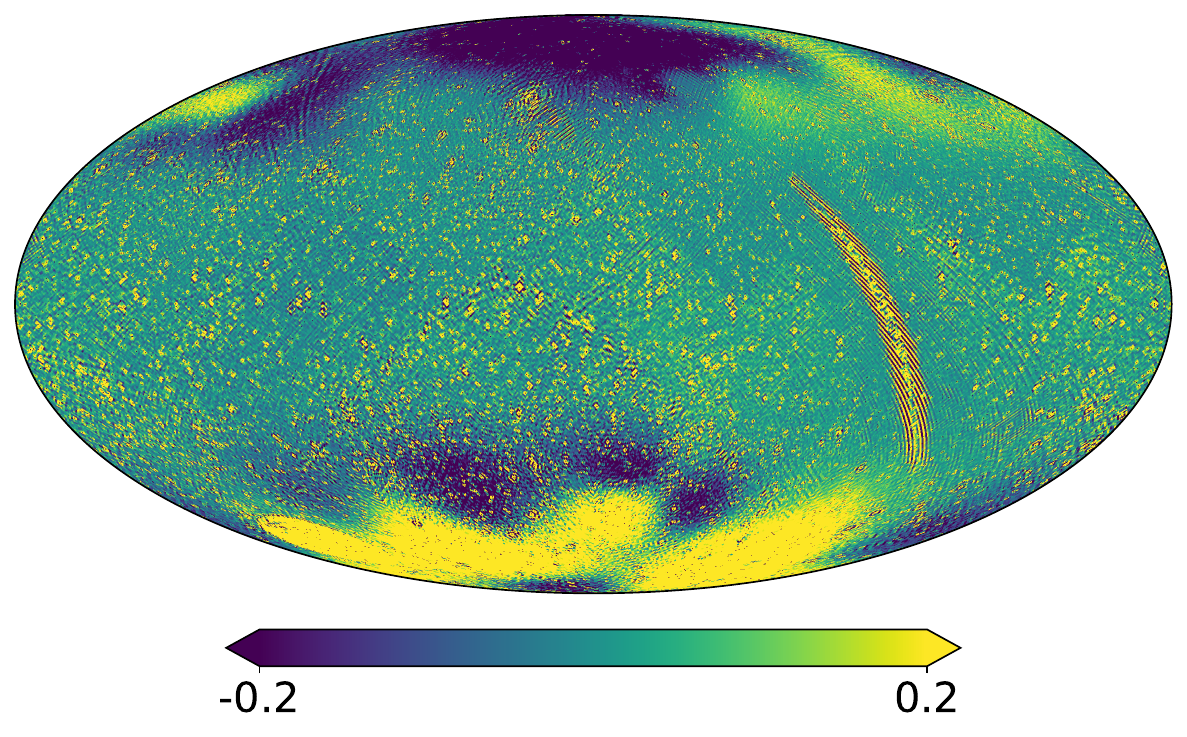}
\includegraphics[width=\columnwidth]{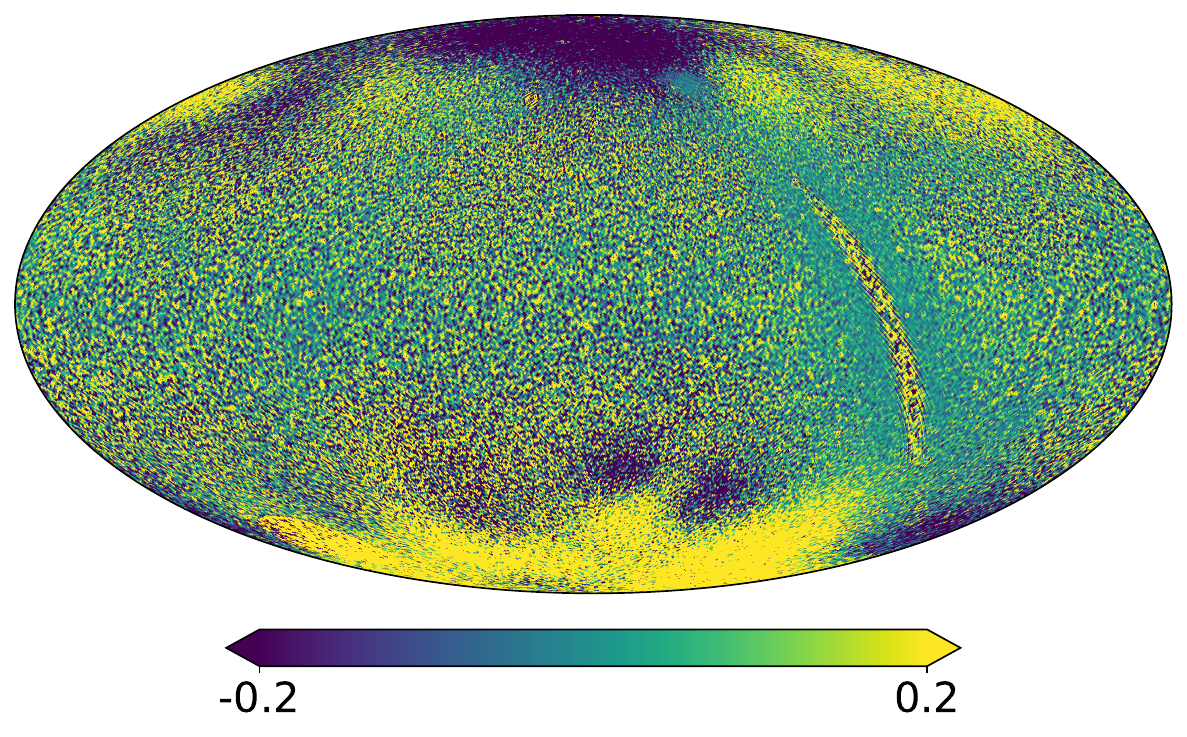}\\
\caption{Same as Figure~\ref{fig:resolution_test_NSIDE256} but at 10.0 MHz.}
\label{fig:freq_test_10MHz}
\end{figure*}

Next, we test \texttt{AMIGO} at 10.0 MHz with NSIDE = 256, where the $uvw$ coverage of the DSL mission is worse. In Figure~\ref{fig:freq_test_10MHz}, we plot the reconstructed maps in a similar way to Figure~\ref{fig:resolution_test_NSIDE256}. 
We find that the fractional errors look much lower than the reconstructed maps with no-prior maps in \citetalias{furen} but higher than those with prior maps. As noted in \citetalias{furen}, the polar regions are not well reconstructed at relatively higher frequencies, due to the lack of short projected baselines for these regions. However, unlike the dark polar caps in \citetalias{furen}, \texttt{AMIGO} can reconstruct the undersampled polar regions reasonably to some extent; the southern polar regions on the reconstructed maps are brighter while the northern polar regions are darker. The poorly reconstructed areas are also smaller compared to the no-prior results in \citetalias{furen}. This is because the angular power spectrum prior compensates for the large-scale modes in a globally uniform way. Therefore, the reconstructed sky temperatures are overestimated in the southern polar regions, where the input temperatures are below the average, but underestimated in the northern polar regions, where the input temperatures are above the average.

Aside from visual comparisons, we also list the normalized Mean Squared Error (MSE) and the Structural
SIMilarity (SSIM) for all cases in Table~\ref{tab:cases}. \revise{Although commonly used in computer vision, SSIM can also quantify the spatial structural consistency between the reconstructed and true skymaps, serving as a complementary metric to MSE, which only evaluates pixel-wise deviations.} The MSE and SSIM are defined as:
\begin{eqnarray}
    {\rm MSE}(I,R) &=& \frac{1}{N_{\rm pix}}\sum_i^{N_{\rm pix}}
    \frac{(I_i-R_i)^2}{\sigma^2(I)}\\
    {\rm SSIM}(I,R) &=& \frac{(2\bar{I}\bar{R}+c_1)[2{\rm Cov}(I,R)+c_2]}{(\bar{I}^2+\bar{R}^2+c_1)[\sigma^2(I)+\sigma^2(R)+c_2]},
\end{eqnarray}
where $I$ and $R$ denote the input and reconstructed maps at the same frequency and resolution respectively, $\bar{I}$ and $\bar{R}$ are the mean values, $\sigma(I)$ and $\sigma(R)$ are the standard deviations, $c_1$ and $c_2$ are small constants to maintain \revise{numerical stability}. Here we set $c_1= (0.01L)^2$ and $c_2= (0.02L)^2$, where $L$ is the range of the input map. 
We find that the imaging quality at 3.0 MHz decreases rapidly with increasing spatial resolution, no matter whether thermal noise is included or not. At 10.0 MHz, the decrease in quality is more gentle, while at 30 MHz, we even obtain improved maps at higher resolutions. This is because, given the same baseline distribution, smaller-scale structures are relatively insufficiently probed at lower frequencies, while large-scale structures are relatively insufficiently probed at higher frequencies.

\begin{table*}[!h]
    \centering
    \caption{Summary of Cases --- \revise{the Frequency, the Spatial Resolution (NSIDE), the Prior Threshold $\Delta^{\rm thre}_H$, the Ratio $C_\ell^{\rm prior}/C_\ell^{\rm inp}$, the Maximum Baseline Length $b_{\rm max}$, Batch Size $N_{\rm mini}$, Batch Ordering, Thermal noise (Y/N), MSE, SSIM, the Prior Residual $\Delta_H$ and $\Delta_G$.}} 
    \begin{tabular}{c|c|c|c|c|c|c|c|c|c|c|c}
    \hline\hline
         Freq [MHz] & NSIDE & $\Delta^{\rm thres}_H$& $C_\ell^{\rm prior}/C_\ell^{\rm inp}$& $b_{\rm max}/b_{\rm NQ}$ &$N_{\rm mini}$& Order & Noise &MSE &SSIM &$\Delta_H$&$\Delta_G$\\
         \hline
         3 & 64 & 0.01 &1.00 & 1 &$2^{18}$ &ascending & yes &0.066 &0.979 &9.98$\times10^{-3}$ &5.25$\times10^{-3}$ \\ 
         3 & 128 & 0.01&1.00 & 1 &$2^{18}$ &ascending & yes &0.191 &0.949 &9.27$\times10^{-3}$ &1.31$\times10^{-3}$\\ 
         3 & 256 & 0.01 &1.00& 1 &$2^{18}$ &ascending & yes &0.546 &0.868 &2.92$\times10^{-3}$ &6.63$\times10^{-6}$\\ 
         3 & 64 & 0.01 &1.00& 1 &$2^{18}$ &ascending & no &0.053 &0.983 &5.18$\times10^{-3}$ &8.57$\times10^{-4}$\\
         3 & 64 & 0.005 &1.00& 1 &$2^{18}$ &ascending & no &0.053 &0.983 &1.40$\times10^{-2}$ &4.19$\times10^{-3}$\\
         3 & 64 & 0.05 &1.00& 1 &$2^{18}$ &ascending & no &0.055 &0.982 &2.21$\times10^{-2}$ &2.16$\times10^{-3}$\\
         3 & 64 & 0.01 &1.05 & 1 &$2^{18}$ &ascending & no &0.053 &0.983 &9.52$\times10^{-3}$ &2.46$\times10^{-3}$ \\ 
         3 & 64 & 0.01 &1.50 & 1 &$2^{18}$ &ascending & no &0.303 &0.944 &8.45$\times10^{-3}$ &$<10^{-7}$ \\ 
         3 & 64 & 0.01 &1.00& 1 &$2^{16}$ &ascending & no &0.054 &0.983 &7.35$\times10^{-3}$ &1.51$\times10^{-3}$\\
         3 & 64 & 0.01 &1.00& 1 &$2^{20}$ &ascending & no &0.053 & 0.983 &9.69$\times10^{-3}$ &8.74$\times10^{-3}$\\
         3 & 64 & 0.01 &1.00& 1 &$2^{18}$ &descending & no&0.054 &0.983 &9.74$\times10^{-3}$ &6.45$\times10^{-4}$\\
         3 & 64 & 0.01 &1.00& 1 &$2^{18}$ &shuffle & no&0.054 &0.983&9.72$\times10^{-3}$ &1.83$\times10^{-3}$\\
         3 & 64 & 0.01 &1.00& 2 &$2^{18}$ &ascending & no&0.245 &0.923 &4.52$\times10^{-3}$ &$<10^{-7}$\\
         3 & 64 & 0.01 &1.00& 4 &$2^{18}$ &ascending & no&0.599&0.807&7.08$\times10^{-3}$ &3.92$\times10^{-4}$\\
         3 & 128 & 0.01 &1.00& 1 &$2^{18}$ &ascending & no&0.158 &0.959 &4.37$\times10^{-3}$ &4.38$\times10^{-4}$ \\ 
         3 & 256 & 0.01 &1.00& 1 &$2^{18}$ &ascending & no&0.497 &0.879 &6.58$\times10^{-4}$ &1.65$\times10^{-5}$\\
         10 & 64 & 0.01 &1.00& 1 &$2^{18}$ &ascending & yes &0.160 &0.963 &3.09$\times10^{-3}$ &1.65$\times10^{-5}$\\
         10 & 128 & 0.01 &1.00& 1 &$2^{18}$  &ascending & yes&0.198 &0.970 &4.44$\times10^{-3}$ &8.54$\times10^{-4}$\\
         10 & 256 & 0.01 &1.00& 1 &$2^{18}$ &ascending & yes&0.354 &0.957 &2.90$\times10^{-3}$ &3.82$\times10^{-3}$\\
         10 & 64 & 0.01 &1.00& 1 &$2^{18}$ &ascending & no&0.086 &0.980 &9.69$\times10^{-3}$ &7.85$\times10^{-4}$\\
         10 & 128 & 0.01 &1.00& 1 &$2^{18}$ &ascending & no&0.123 &0.981 &6.96$\times10^{-4}$ &$<10^{-7}$\\
         10 & 256 & 0.01 &1.00& 1 &$2^{18}$ &ascending & no&0.295 &0.964 &4.43$\times10^{-4}$ &5.27$\times10^{-7}$\\
         30 & 64 & 0.03 &1.00& 1 &$2^{18}$ &ascending & yes&0.393 &0.898 &2.37$\times10^{-2}$  &2.69$\times10^{-3}$\\
         30 & 128 & 0.03 &1.00& 1 &$2^{18}$ &ascending & yes&0.347 &0.947 &2.49$\times10^{-2}$ &2.30$\times10^{-3}$\\
         30 & 256 & 0.03 &1.00& 1 &$2^{18}$ &ascending & yes&0.367 &0.960 &1.85$\times10^{-2}$ &4.22$\times10^{-3}$\\
         30 & 64 & 0.03 &1.00& 1 &$2^{18}$ &ascending & no&0.242 &0.943 &2.96$\times10^{-3}$ &$<10^{-7}$\\
         30 & 128 & 0.03 &1.00& 1 &$2^{18}$ &ascending & no&0.172 &0.974 &2.78$\times10^{-2}$ &1.03$\times10^{-3}$\\
         30 & 256 & 0.03 &1.00& 1 &$2^{18}$ &ascending & no&0.239 &0.974 &1.82$\times10^{-2}$ &7.93$\times10^{-3}$\\
         \hline
    \end{tabular}
    \label{tab:cases}
\end{table*}

\section{Discussion}
\label{sec:discuss}
In this section, we discuss the impact of various technical setups, including prior control and batch configuration. \revise{We also verify the convergence behavior with respect to initial guesses and noise realizations} \revise{, and} compare our method with the Tikhonov Regularization approach in \citetalias{furen} \revise{in terms of computational cost and imaging quality}. In addition to MSE and SSIM listed in Table~\ref{tab:cases}, we also quantify the reconstruction error with the correlation coefficient between the input and reconstructed maps, which is given by
\begin{eqnarray}
    \rho_\ell = \frac{C^X_\ell}{\sqrt{C_\ell^{\rm inp}C_\ell^{\rm recon}}}\,,
\end{eqnarray}
where $C_\ell^{\rm inp}$, $C_\ell^{\rm recon}$, and $C^X_\ell$ are the auto angular power spectrum of the input map, that of the reconstructed map, and the cross angular power spectrum between the two, respectively.

\subsection{Prior Control}
\label{subsec:priortest}
\begin{figure*}[!h]
\centering
\includegraphics[width=0.64\columnwidth]{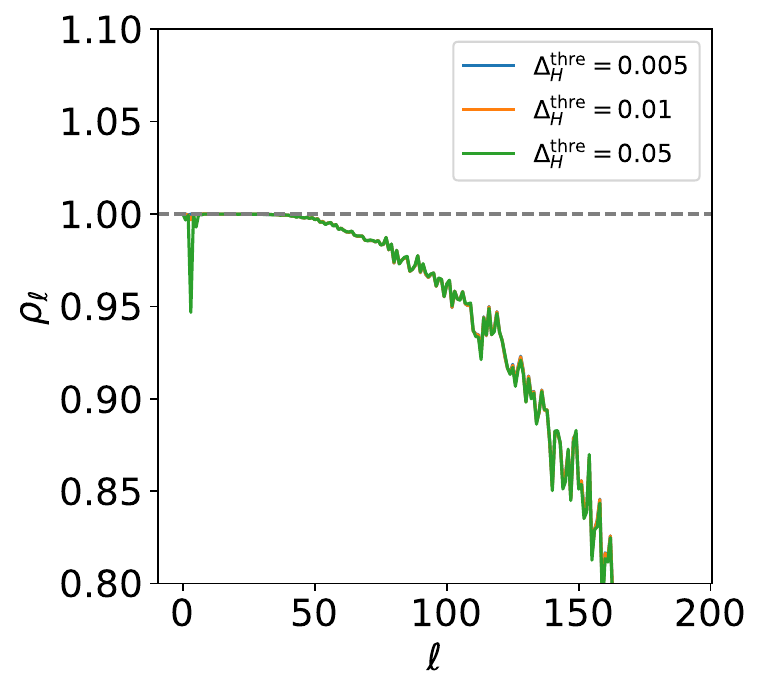}
\includegraphics[width=0.64\columnwidth]{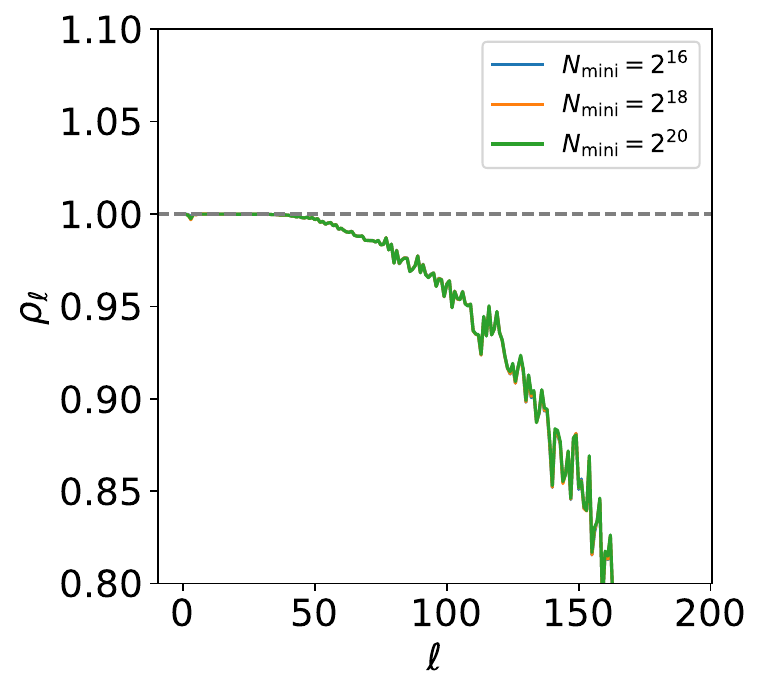}
\includegraphics[width=0.64\columnwidth]{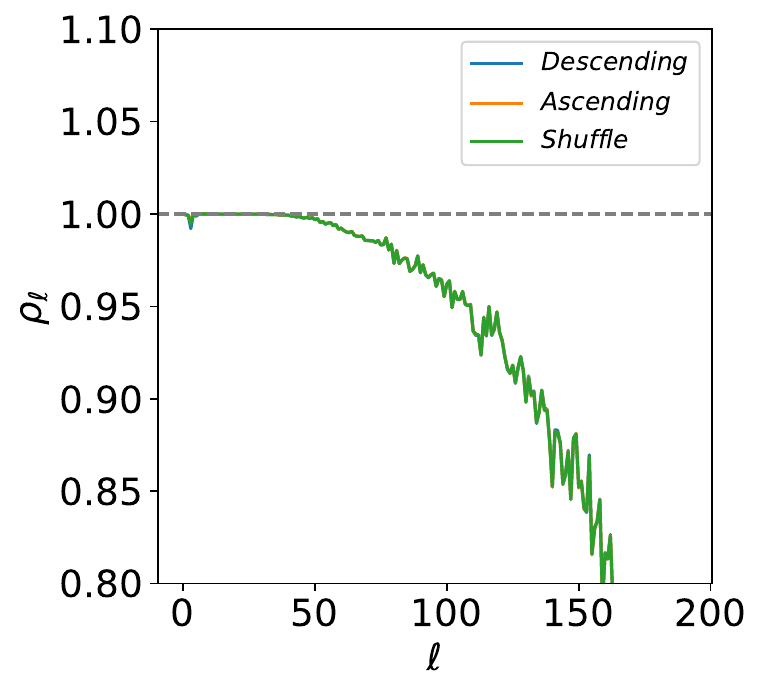}

\caption{$\rho_\ell$ over various prior settings and batch configurations. We vary the strength of the prior in the left panel, the mini-batch size in the middle, and the ordering in the right. We find that $\rho_\ell$ drops at $\ell\sim3$ with $\Delta_H^{\rm thre}=0.05$. We do not find significant differences among the batch configurations tested in this work.}
\label{fig:configuration_test}
\end{figure*}

We first explore the impact of prior strength on imaging quality. The constraint from $G$, i.e., non-negative temperature prior, should always be a strong one. Therefore, we fix $\Delta^{\rm thres}_G = $ 0.01 in all cases and only vary $\Delta_H^{\rm thres}$. In principle, one should get a reconstructed map generically closer to the prior with a tight $\Delta_H^{\rm thres}$. We plot the correlation coefficient in the left panel of Figure~\ref{fig:configuration_test}. We do not notice a significant decrease in MSE or SSIM with a loose threshold ($\Delta_H^{\rm thres}=0.05$), but do find that $\rho_\ell$ drops at $\ell\sim3$, indicating that structures at this scale are not well reconstructed \revise{and that these modes should be prior-dominated}. This is consistent with a relatively weak prior of the all-sky angular power spectrum. \rerevise{The identification of prior-dominated modes around $\ell\sim3$ is specific to the orbit configuration and the observing strategy adopted in this work, which should not be interpreted as a universal property of the algorithm itself. In other configurations with different {\it uvw} coverage or baseline distributions, the prior-dominated scales may differ.}

In the above tests, we use the correct $C_\ell^{\rm inp}$ as the prior. However, in practice, the true $C_\ell^{\rm prior}$ is unknown. 
\revise{To assess how sensitive the reconstruction is to different levels of prior mismatches,} we adjust the true prior by multiplying it with factors of 1.05 and 1.50, respectively, and then reconstruct sky maps at 3.0 MHz with NSIDE=64 and $\Delta_H^{\rm thres}=0.01$. The former represents a case where the errors in the prior are tolerable within statistical uncertainty (small error case), while the latter represents a case where the adopted prior contains some substantial errors (large error case). 

\begin{figure*}[!ht]
    \centering
    \includegraphics[width=0.64\columnwidth]{recon_noisefree_freq_3.0_nside_64.pdf}
    \includegraphics[width=0.64\columnwidth]{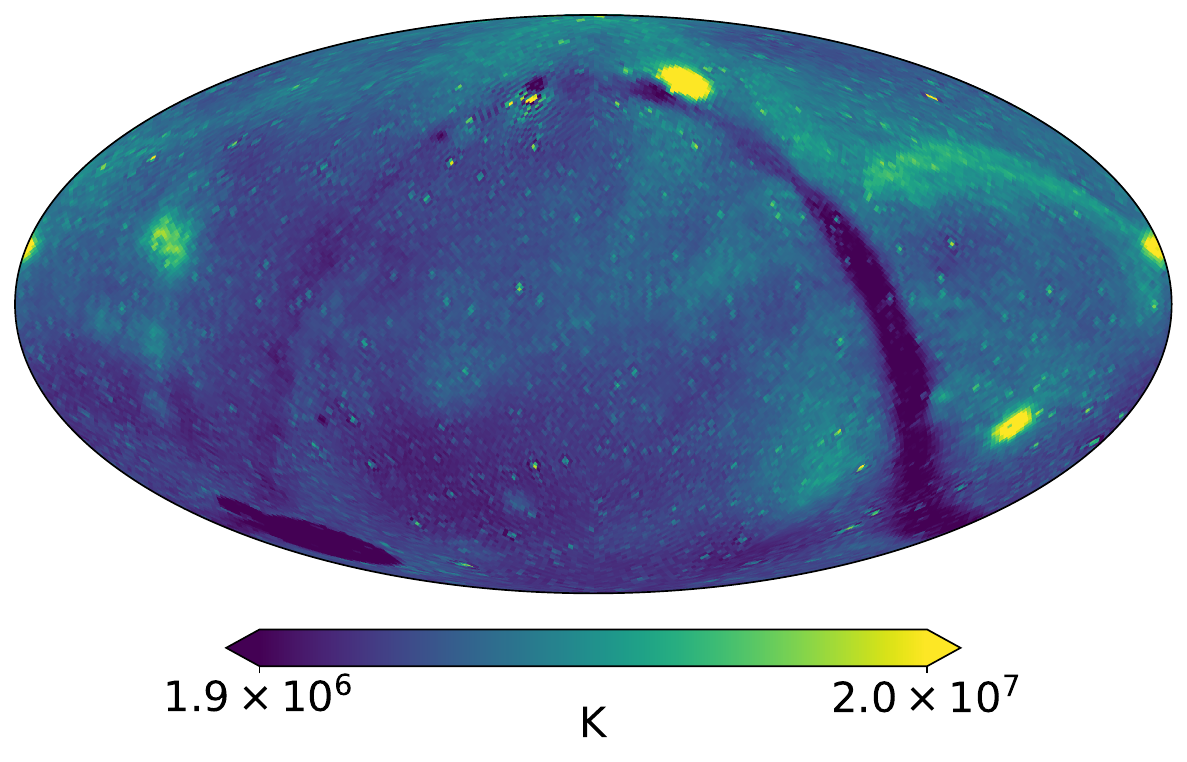}
    \includegraphics[width=0.64\columnwidth]{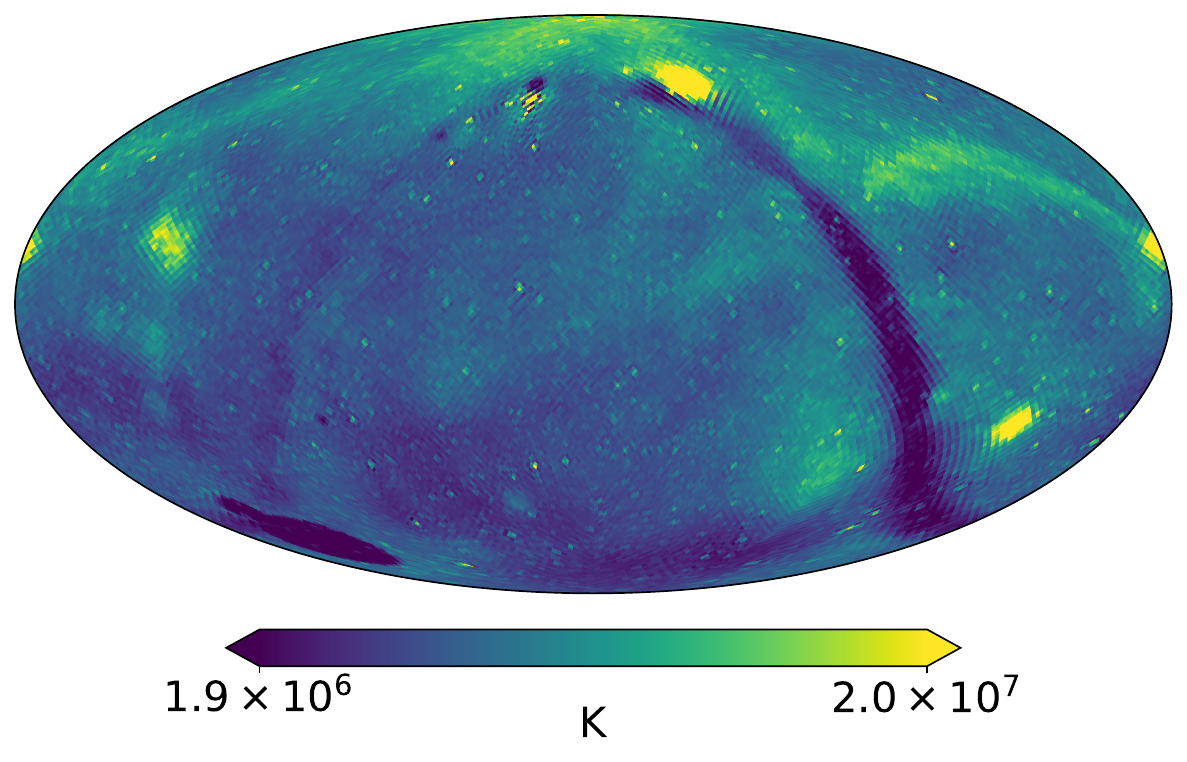}
    \caption{The reconstructed maps at 3.0 MHz with NSIDE = 64, using different $C_\ell^{\rm prior}$. We use $1\times C_\ell^{\rm inp}$ as the prior in the left panel, $1.05\times C_\ell^{\rm inp}$ in the middle, and $1.50\times C_\ell^{\rm inp}$ in the right. The reconstructed map in the large error case appears much brighter and displays artificial stripes.}
\label{fig:incorrect_prior}
\end{figure*}

We plot the reconstructed maps in Figure~\ref{fig:incorrect_prior} along with the map obtained with the correct $C_\ell^{\rm prior}$. In the small error case, the resulting errors in the reconstructed map are acceptable. In the large error case, however, the reconstructed map appears significantly brighter and displays artificial stripes, \revise{with notable degradations} in both MSE and SSIM. \revise{The results with moderate mismatches should fall between these two cases.}

In practice, it may be difficult to assess the downgradation of imaging quality without knowing the true sky map. However, we can still be aware that the adopted prior contains some substantial errors from the unacceptable number of iteration steps required to meet the same stopping criteria. For example, it takes 734 iterations to stop in the small error case above, while 5424 iterations are required in the large error case. \revise{In such cases, one may} either choose a loose threshold, which lowers the prior strength, or try to \revise{correct} the prior by comparing it with $C_\ell$ derived from a candidate reconstructed map at a particular iteration step. 



\subsection{Batch Configuration}
\label{subsec:batchtest}
Next, we vary the batch configuration and explore its impact on the imaging quality. There are two axes about batch configuration: ordering and size. In machine learning, training may benefit a lot from shuffling the order of data and using a small batch size of $2^5$ \citep{2018arXiv180407612M}. However, for interferometric imaging, it is not necessary to follow both traditions. One may obtain better imaging quality using visibility data in a certain order because they anticipate image reconstruction at different scales. Additionally, since we do not grid the visibility data in {\it{uvw}} space, the thermal noise level on a single visibility function can be high. One needs to load visibility data of baselines redundant in time \revise{into a single} mini-batch to suppress the thermal noise level. 

In order to explore the impact of batch size, we arrange visibility data batches from short to long baselines, i.e., an ascending ordering, and reconstruct maps with $N_{\rm mini}= 2^{16}$, $2^{18}$, and $2^{20}$, respectively. We list the MSE and SSIM of the reconstructed maps in Table~\ref{tab:cases}, and plot $\rho_\ell$ in the middle panel of Figure~\ref{fig:configuration_test}. We do not find significant differences in MSE, SSIM, \revise{or} $\rho_\ell$, \revise{indicating that} the imaging quality is insensitive to $N_{\rm mini}$ over $2^{16}$.

Then we try different orderings of mini-batches. We fix $N_{\rm mini}=2^{18}$ and try a traditional shuffle ordering, i.e., randomly shuffling visibility data before each epoch, and a descending ordering, i.e., arranging visibility data from long baselines to short baselines. We also list the MSE and SSIM in Table~\ref{tab:cases}, and plot $\rho_\ell$ in the right panel of Figure~\ref{fig:configuration_test}. Again, we do not find significant differences here, indicating that the imaging quality is insensitive to ordering. However, we still recommend the ascending order \revise{as shuffling incurs additional computational overhead and the descending order requires more iteration steps.}

\subsection{\revise{Convergence Verification}}
\label{subsec:convergence}
\revise{As an iterative algorithm, it is important to ensure that {\tt AMIGO} converges reliably with respect to initial guesses and noise realizations. Although the cost function is globally convex, the nonlinear terms and incomplete {\it uvw} coverage may introduce local minima, resulting in some dependence on the initial guess and the noise realization.}

\begin{figure*}[!ht]
    \centering
    \includegraphics[ width=0.64\columnwidth,valign=t]{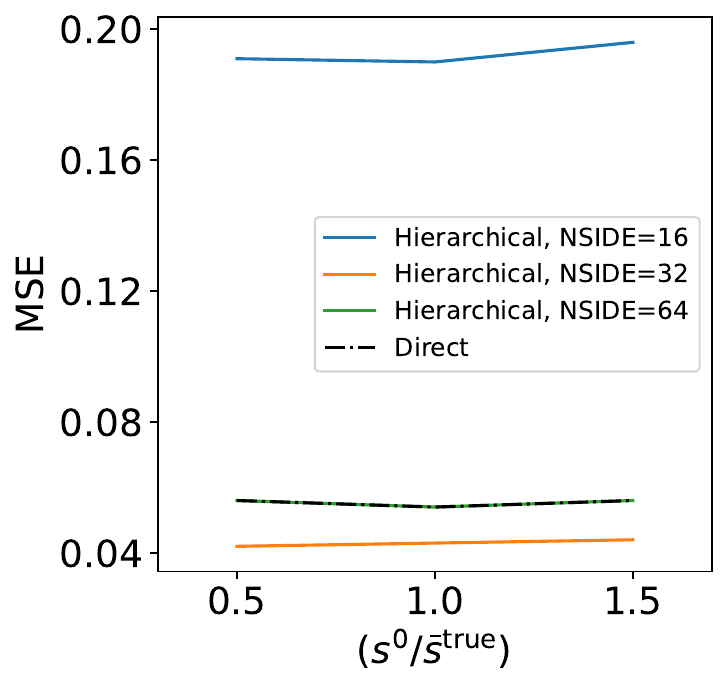}
    \includegraphics[ width=0.64\columnwidth, valign=t]{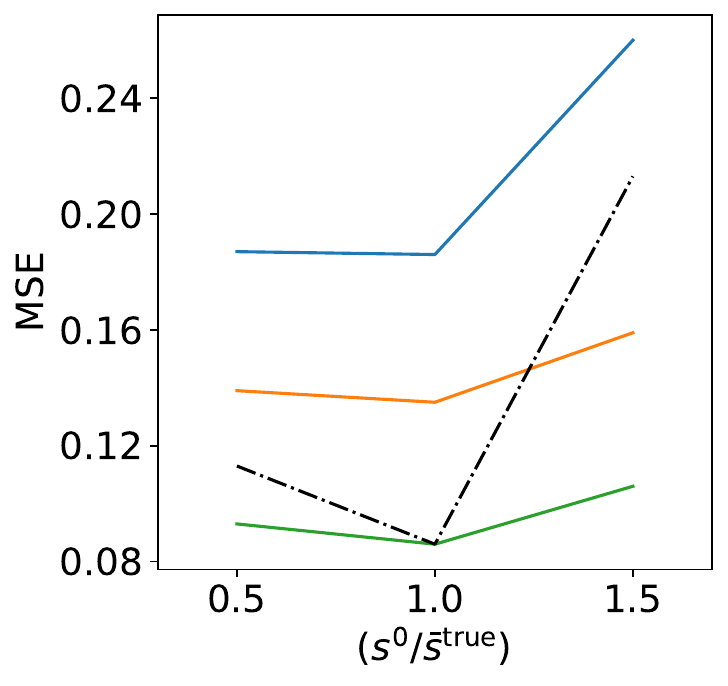}
    \includegraphics[ width=0.65\columnwidth, valign=t]{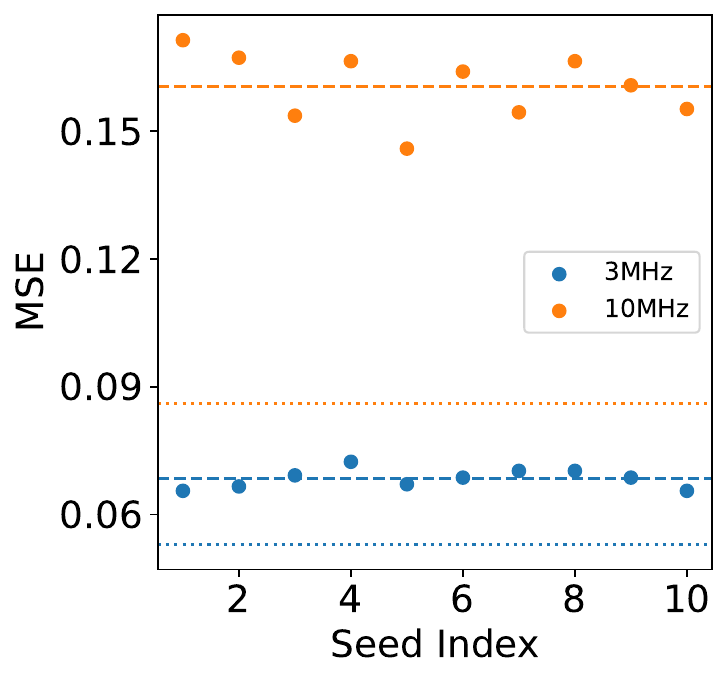}
    \caption{\revise{Convergence verification of {\tt AMIGO} with respect to the initial guesses and the noise realizations. We plot the MSE of reconstructed maps at 3.0 MHz (Left) and 10.0 MHz (Middle) as a function of the ratio ($s^0/\bar{s}^{\rm true}$). Unbiased flat initial guesses consistently yield good reconstructions. The hierarchical setting can alleviate dependence on biased initial guesses. In the right panel, we plot the MSE for various noise realizations. The convergence behavior against noise realizations is preserved within 10\% of the averaged level indicated by the dashed lines, and far above the noiseless level indicated by the dotted lines of the corresponding colors.}}
    \label{fig:convergence_initial}
\end{figure*}

\revise{We anticipate that the hierarchical initial guess can mitigate this dependence, as it enables the reconstruction to circumvent part of the local minima. We have shown that {\tt AMIGO} can produce well-reconstructed maps using the hierarchical initial guess of unbiased $s^0$ from NSIDE = 64 to 256. Here, we verify the convergence behavior of {\tt AMIGO} with biased initial values, i.e., $s^0 \neq \bar{s}^{\rm true} $. Without loss of generality, we start from NSIDE = 16 with a constant value of $s^0$, then upgrade to NSIDE = 32 and 64 hierarchically. To demonstrate that the hierarchical initial guess facilitates convergence, we compare the results with maps obtained by starting directly from flat initial guesses of biased $s^0$ at NSIDE = 64. We plot the MSE of reconstructed maps at 3.0 MHz and 10.0 MHz as a function of the ratio $s^0/\bar{s}^{\rm true}$ in the left and middle panels of Figure~\ref{fig:convergence_initial}, respectively. We notice that unbiased initial values $s^0$ consistently yield well-reconstructed maps regardless of the initialization strategy. At 3.0 MHz, with nearly complete {\it uvw} coverage and fewer local minima, {\tt AMIGO} is insensitive to $s^0$, regardless of whether initial guesses are set hierarchically. By contrast, the dependence on $s^0$ gets stronger at 10.0 MHz because of the increasing number of local minima arising from incomplete {\it uvw} coverage and nonlinear terms in the cost function. The hierarchical initial guess alleviates this dependence and still produces stable reconstructions. Therefore, for the rest of this paper, we only employ unbiased $s^0$.}

\revise{We also validate the convergence behavior of {\tt AMIGO} against noise realizations. We vary the random seeds used to generate thermal noise and reconstruct the sky maps, starting from an unbiased initial guess at NSIDE = 64. In the right panel of Figure~\ref{fig:convergence_initial}, we plot the MSE for various random seeds at 3.0 MHz and 10.0 MHz. For clarity, we also plot the averaged MSE level with dashed lines and the noiseless MSE level with dotted lines. We find that convergence against noise realizations is well preserved at 3.0 MHz. The scatter of the MSE values increases modestly at 10.0 MHz but still remains within 10\% of the averaged level and far above the noiseless level.}

\subsection{\revise{Comparison with \citetalias{furen} : Computational Cost}}

\revise{In this subsection, we compare {\texttt{AMIGO}} with the Tikhonov Regularization approach (TR) in \citetalias{furen} in terms of computational cost. We have already shown that {\tt AMIGO} can successfully reconstruct maps with NSIDE $=$ 256, while the computational cost of the TR approach is unaffordable with NSIDE $>$ 64. In order to provide concrete computational benchmarks, we apply both methods to the same realization of visibility and baseline data at 1 MHz to reconstruct skymaps at different resolutions. We use an NVIDIA RTX 3090 GPU to run {\texttt{AMIGO}} and to compute the dirty maps and beams for TR. The clean maps of TR, i.e., the deconvolved maps, are post-processed using 20 cores of an AMD EPYC 9654 CPU. For simplicity, we do not include the prior map correction for TR in this comparison.}

\revise{We increase the number of pixels to be reconstructed, and the number of data to be loaded coordinately, to compare their scaling relations. The configurations and computational costs for all cases are listed in Table~\ref{cases4scaling}. For {\tt AMIGO}, we adopt the hierarchical initial guesses as described in Section~\ref{sec:results} and reconstruct maps with NSIDE ranging from 16 to 64. We find that, for the same amount of data, {\tt AMIGO} requires less memory and wall-clock time than TR, once the maps are reconstructed with NSIDE $\geq$ 32.} \rerevise{We note that the reported absolute wall-clock times are dependent on the specific hardware and implementation details, so they should be regarded primarily as representative benchmarks. The scaling relations constitute the more general conclusions.}

\begin{figure}[!ht]
    \centering
    \includegraphics[width=0.9\columnwidth]{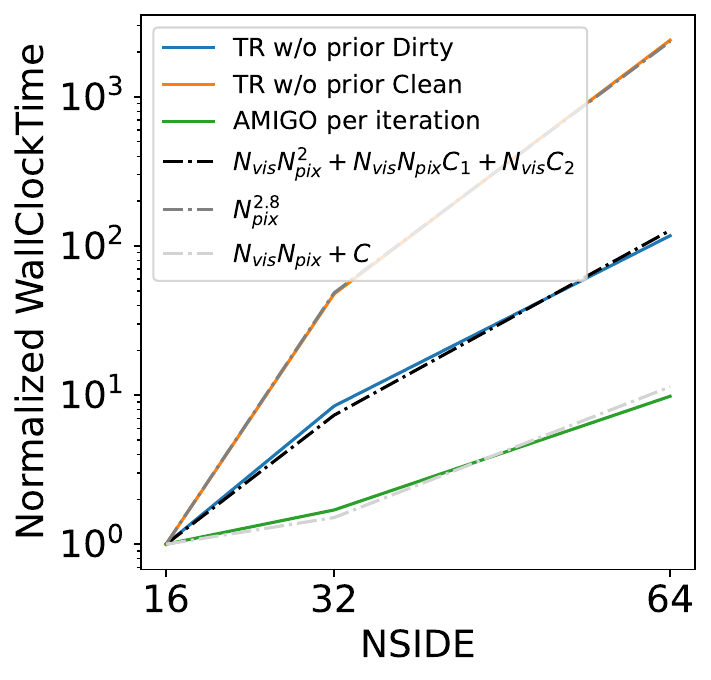}
    \includegraphics[width=0.9\columnwidth]{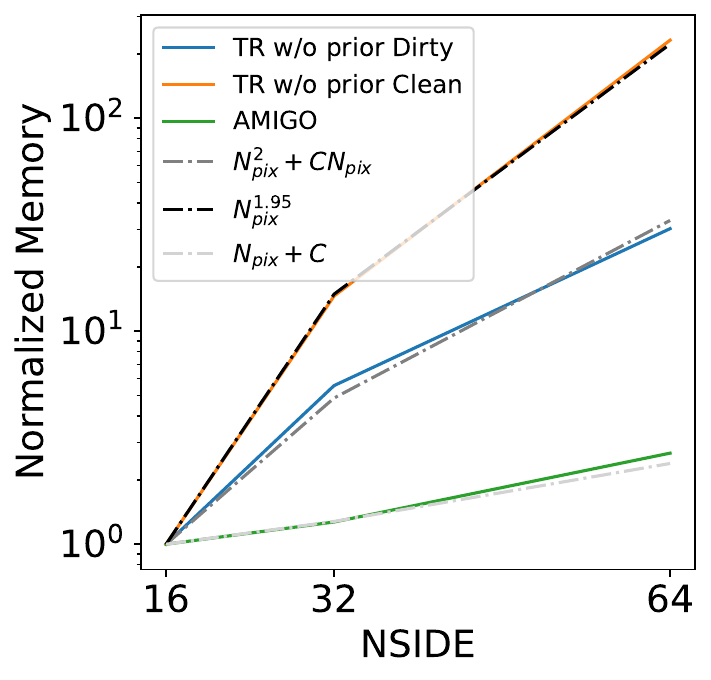}
    \caption{\revise{Computational costs of \citetalias{furen} and {\texttt{AMIGO}} normalized with the minimum case as a function of NSIDE. We also plot the fitted scaling with expected templates in dashed lines. We find that both in time and space, {\tt AMIGO} costs much less and scales much better.}}
    \label{fig:comparison_cost}
\end{figure}

\revise{To better illustrate the scaling relations, we plot the computational costs normalized with the minimum case in Figure~\ref{fig:comparison_cost} as a function of NSIDE. For the TR approach, we plot the costs of dirty map and beam construction and of deconvolution to clean maps separately, since they are performed on different hardware. For {\tt AMIGO}, although the scaling of $N_{\rm itr}$ is somewhat uncertain, we can at least verify that its wall-clock time per iteration scales as $O(N_{\rm vis}N_{\rm pix})$. With iteration counts lower than $N_{\rm pix}$, the total cost should then be below $O(N_{\rm vis}N_{\rm pix}^2)$ as discussed in Section~\ref{sec:method}. We also fit the measured scalings with expected templates\footnote{\revise{For the cleaning in the TR approach, the measured scaling of LAPACK eigensolver for Hermitian matrices has been found slightly better than the theoretical one \citep{demmel2007performance,martinez2025cache}.}}, assuming the initialization overhead and additional cost of constructing the response matrix ${\bf B}$. We find that in both time and space, the scalings of {\tt AMIGO} can be well fitted with the expected templates. These results confirm that {\tt AMIGO} reduces the computational cost relative to the TR approach in \citetalias{furen}, which is its primary advantage.}

\begin{table*}[!ht]
    \centering
    \caption{\revise{Comparison of computational costs between \citetalias{furen} and {\texttt{AMIGO}} --- Method, NSIDE, Number of Pixels $N_{\rm pix}$, Number of Loaded Data $N_{\rm data}$ (including visibility and baseline vector), Memory, Wall-Clock Time, Number of Iterations $N_{\rm itr}$}}
    \begin{tabular}{c|c|c|c|c|c|c}
    \hline\hline
      Method & NSIDE & $N_{\rm pix}$ & $N_{\rm data}$ &  Memory [GB] & Wall-clock Time [h] & $N_{\rm itr}$\\\hline
        TR w/o prior & 16 & 3072 & 3.8e5 & 0.99/0.155  & 0.0213/0.0005 & N/A\\
        TR w/o prior & 32 & 12288 & 1.8e6 & 5.5/2.267  & 0.18/0.0238 & N/A\\
        TR w/o prior & 64 & 49152 & 8.7e6 & 30.0/36.030  & 2.5/1.2& N/A\\
        AMIGO & 16 & 3072 & 3.8e5 & 0.494  & 0.0710 & 1566\\
        AMIGO & 32 & 12288 & 1.8e6 & 0.628  & +0.0708 & +278\\
        AMIGO & 64 & 49152 & 8.7e6 & 1.322  & +0.829 & +337\\\hline
    \end{tabular}
    \flushleft
    \tablenotetext{}{\revise{Note. --- Here we list two numbers in memory usage and wall-clock time for all TR cases. The first ones are for getting dirty maps and beams, while the second ones are for deconvolving to clean maps.}}
    \label{cases4scaling}
\end{table*}

\subsection{Comparison with \citetalias{furen} \revise{: Imaging Quality}}
\label{subsec:comparison}

In this subsection, we compare {\texttt{AMIGO}} with the \revise{TR} approach in \citetalias{furen} \revise{in terms of imaging quality}. 
\revise{The two methods are not fully equivalent in their strategies for handling aliasing. \citetalias{furen} adopts a pixel-averaging strategy at the level of the response matrix ${\bf B}$, which is evaluated at a higher resolution and then averaged to the target resolution, whereas {\tt AMIGO} just restricts the baselines to the Nyquist limit and evaluates ${\bf B}$ at the target resolution. However, we aim at a {\it fair} comparison here, i.e., one with the same input visibility data and the same target resolution. Therefore, if the baseline range is beyond the Nyquist limit of the target resolution, we use {\tt AMIGO} to reconstruct the sky map at a higher spatial resolution and then downgrade it to the target resolution, which can be considered as a pixel-averaging strategy at the level of the reconstructed map. We also verify that {\tt AMIGO} cannot work beyond the Nyquist limit in Appendix~\ref{app:aliasing}.}

We compare the results without thermal noise at NSIDE = 64 using baselines with $b<b_{\rm NQ}$ and $b<4b_{\rm NQ}$, respectively. On the side of the TR approach, we use the recommended regularization parameter\footnote{The values of MSE, SSIM and $\rho_\ell$ may have minor variations with $\epsilon$, yet such changes will not impact the main conclusions drawn here.} in \citetalias{furen}, i.e., $\epsilon=10^{-6}$ for 3.0\,MHz and $\epsilon=10^{-4}$ for 10.0\,MHz. Besides the no-prior reconstruction, we also employ a prior map correction using the input map smoothed with a Gaussian filter of FWHM=$5^\circ$. \revise{For {\texttt{AMIGO}}, we reconstruct the skymap at NSIDE = 64 and 256, using baselines with $b<b_{\rm NQ}$ and $b<4\,b_{\rm NQ}$, respectively. Then we downgrade the high-resolution map (NSIDE = 256) to the target resolution (NSIDE = 64).}

\begin{figure}[h!]
    \centering
\includegraphics[width=0.9\columnwidth]{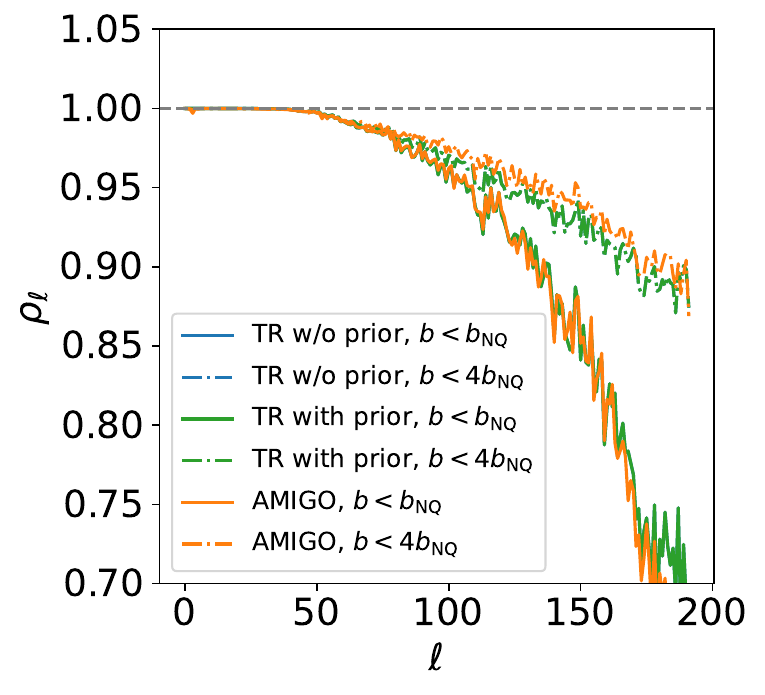}
\includegraphics[width=0.9\columnwidth]{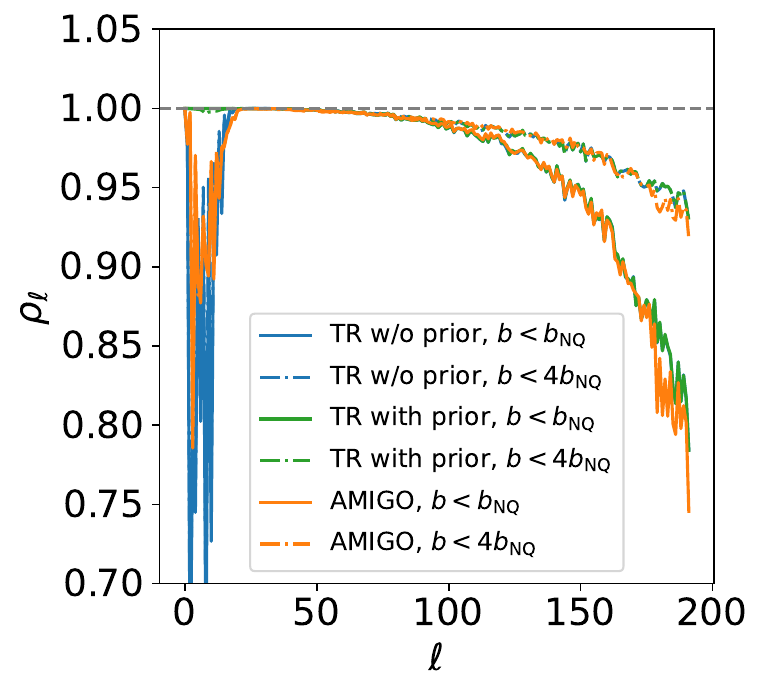}

\caption{Comparison of $\rho_\ell$ between \citetalias{furen} and {\texttt{AMIGO}}. We plot $\rho_\ell$ at 3.0\,MHz in the \revise{top} panel and 10.0\,MHz in the \revise{bottom}, respectively. We find that with a {\it correct} prior map, TR is better than {\texttt{AMIGO}} at 10.0\,MHz but does not have significant improvements at 3.0\,MHz. \revise{The fractional differences between the reconstructed maps at 3.0\,MHz with and without the prior correction are around $10^{-5}$, unnoticeably small in the $\rho_\ell$  plot in the top.}}
\label{fig:comparison_nq}
\end{figure}

We list the values of MSE and SSIM in Table~\ref{tab:comparison} and plot $\rho_\ell$ in Figure~\ref{fig:comparison_nq} for both methods. We find that with the same range of baselines, {\texttt{AMIGO}} outperforms TR in the absence of a prior map. \revise{In principle, more accurate prior information should yield better imaging quality. A prior angular power spectrum provides more information than having no prior, leading to better image reconstructions.} When a  {\it correct} prior map is available, TR is better than {\texttt{AMIGO}} at 10.0\,MHz. \revise{Similarly, this is because a prior map provides more information than a prior angular power spectrum.} \revise{However, the prior map correction does not have significant improvements at 3.0 MHz.}
\revise{In this case, the supplementary information from the {\it smoothed} input map is likely less than that from the true angular power spectrum because the {\it uvw} coverage at 3.0 MHz is relatively complete.} 
\revise{We also note that the most substantial improvement of {\tt AMIGO} over TR without prior map correction is around $\ell\sim 3$, which confirms that these modes are prior-dominated.} Additionally, we find that the reconstructed maps using baselines with $b<4\,b_{\rm NQ}$ are consistently better than those using only baselines with $b<b_{\rm NQ}$. \revise{The improvements from longer 3D baselines are comparable between the two methods, indicating that the pixel-averaging at the level of the response matrix and at the level of the reconstructed map are roughly equivalent.} Therefore, one can use a pixel-averaging strategy to incorporate information from longer \revise{3D} baselines and improve imaging quality.

\revise{Note that here the comparison is performed with {\it correct} priors. Strong bias has been observed in Section~\ref{subsec:priortest} and in \citetalias{furen} when there are significant errors in the priors. The comparison results may depend on the accuracy of the adopted priors. However, this is not a major concern, since we have shown in Section~\ref{subsec:priortest} that with {\tt AMIGO}, substantial errors can at least be detected from the unacceptable number of iteration steps.}


\begin{table}[!h]
    \centering
    \caption{Comparison with \citetalias{furen} --- the Frequency, the Method, the Maximum Baseline Length $b_{\rm max}$, MSE, and SSIM}
    \begin{tabular}{c|c|c|c|c}
    \hline\hline
         Freq [MHz] & Method & $b_{\rm max}/b_{\rm NQ}$& MSE & SSIM\\
         \hline
         3 & TR w/o prior & 1 & 0.116 & 0.966\\ 
         3  & TR w/o prior & 4& 0.034 & 0.989\\ 
         3 & TR with prior & 1 & 0.116 & 0.966\\ 
          3  & TR with prior &4& 0.034 & 0.989\\ 
          3 & \texttt{AMIGO} & 1 & 0.054 &0.983\\
          3 & \texttt{AMIGO} & 4 & 0.021 &0.993\\ 
         10 & TR w/o prior & 1 & 0.668 & 0.877 \\
         10  & TR w/o prior &4& 0.580 & 0.892 \\
         10 & TR with prior & 1 & 0.053 & 0.988\\
         10  & TR with prior &4& 0.020 & 0.995\\
         10  & \texttt{AMIGO} & 1&  0.086 &0.980\\
         10 & \texttt{AMIGO} &4& 0.064 & 0.985\\ 
         \hline
    \end{tabular}
    \label{tab:comparison}
\end{table}

\section{Conclusion}
\label{sec:conclusions}
We have developed a novel synthesis imaging algorithm {\texttt{AMIGO}}, integrating the Mini-Batch Gradient Descent (MBGD) and the Augmented Lagrangian Multiplier (ALM) technique to address the challenges of all-sky radio interferometric imaging with lunar-orbiting arrays like the DSL mission. Compared to traditional matrix inversion-based methods, {\texttt{AMIGO}} reduces the memory scaling from $O(N^2)$ to $O(N)$ and the time scaling from $O(N^3+MN^2)$ to $O(<MN^2)$. \revise{This computational advantage of {\tt AMIGO} makes it feasible to process the large datasets expected from the DSL mission, which is its primary advantage.} Meanwhile, ALM effectively incorporates \revise{physical priors, including} a non-negative sky temperature constraint and a prior angular power spectrum constraint that compensates for large-scale information loss from incomplete $uvw$ coverage. \revise{The general framework is also flexible enough to accommodate additional constraints.} Adjustable stopping criteria further allow quantitative control of prior strength, balancing data fidelity and prior knowledge.  

We use mock visibility data with one promising configuration of the DSL mission to \revise{validate the performance} of this algorithm across frequencies and spatial resolutions. At NSIDE = 64 and 3.0 MHz, fractional errors remain within 20\%, with a normalized MSE of 0.053 and SSIM of 0.983 when thermal noise is excluded; even at NSIDE = 256, the algorithm retains structural \revise{consistency} for diffuse sky features and point sources, though MSE increases moderately. 

\revise{We confirm that {\tt AMIGO} is capable of controlling the strength of prior and notice degraded reconstruction of large-scale structures at $\ell \sim 3$ with a loose threshold $\Delta_{H}^{\rm thre} = 0.05$ at 3.0 MHz, which indicates the prior-dominated scales.} The algorithm also demonstrates strong robustness to batch configuration: varying the batch size or ordering has no significant impact on MSE, SSIM, or $\rho_{\ell}$.

\revise{Comparisons with the TR approach in \citetalias{furen} quantitatively verify the computation gain of {\tt AMIGO} through benchmark computations on fixed hardware with measured wall-clock time and memory scaling. In terms of imaging quality, when using visibility data of the same baseline ranges, {\tt AMIGO} outperforms in scenarios where there is no prior map correction or the smoothed prior map cannot provide useful information.}

\revise{We note that \rerevise{the present validation should still be regarded as a proof of concept based on simulated observations, as }all tests in this work rely on mock visibility data with or without thermal Gaussian noise. Real lunar-orbit interferometric observations contain multiple additional systematics, including calibration errors, baseline uncertainties, and beam modeling uncertainties. Specifically, signal degradation and image distortion due to phase errors from baseline determination and time synchronization have been discussed in \citet{paperII}. Simultaneously incorporating all these instrumental effects is beyond the scope of this work.} 
\rerevise{A full end-to-end assessment incorporating realistic calibration pipelines and all instrumental systematics remains an important direction for future work.}

Overall, {\tt AMIGO}\footnote{\revise{The code can be available from the corresponding author upon reasonable request.}} provides \revise{a computationally feasible approach} to all-sky imaging with lunar-orbiting radio interferometers like DSL. Future work could combine other priors within its framework to further enhance performance.

\begin{acknowledgments}
\revise{We thank Wenlin Tang, Chen Gao, Yulun Li, and Weijie Liu for helpful discussion.} This work is supported by the National Key R\&D Program of China No. 2022YFF0504300, China's Space Origins Exploration Program No. GJ11010401 and GJ11010405, the NSFC International (Regional) Cooperation and Exchange Project No. 12361141814, the NSFC Innovation Group Grant 12421003, and by the Specialized Research Fund for State Key Laboratory of Radio Astronomy and Technology.
\end{acknowledgments}
%



\appendix

\section{Gradients \revise{contributed from} angular power spectrum}
\label{app:gradients}
In this section, we derive the gradients of $J^{\rm tot}$ in Eq.~(\ref{eqn:model}) with respect to the sky temperature $s_n$,\revise{
\begin{eqnarray}\label{eqn:gradient}
    \frac{\partial J^{\rm tot}}{\partial s_n} = \sum_g(\frac{\Delta V_g^*}{2\sigma_g^2}B_{gn}+\frac{\Delta V_g}{2\sigma_g^2}B^*_{gn})+\rho_1\sum_\ell(H_\ell+\frac{\lambda_\ell}{\rho_1})\frac{\partial H_\ell}{\partial s_n}+\rho_2 {\rm min}(s_n+\frac{\mu_n}{\rho_2},0)\,,
\end{eqnarray}
where $\Delta V_g=V_g^{\rm sim}-V_g^{\rm obs}$ and $B_{gn}$ is the complex beam response of $V_g^{\rm sim}$ with respect to $s_n$.}
The contributions from the first and the last terms are straightforward to compute. The second term in Equation~(\ref{eqn:gradient}) can be evaluated via \revise{the chain rule and spherical harmonic transform,
\begin{eqnarray}
    \rho_1\sum_\ell(H_\ell+\frac{\lambda_\ell}{\rho_1})\frac{\partial H_\ell}{\partial s_n} &=& \sum_\ell A_\ell\frac{\partial C_\ell^{\rm sim}}{\partial s_n}= \sum_\ell\frac{A_\ell}{2\ell+1}\sum_{m=-\ell}^\ell (\frac{\partial a^*_{lm}}{\partial s_n}a_{lm}+a^*_{lm}\frac{\partial a_{lm}}{\partial s_n})\nonumber\\ &=&\sum_\ell \sum_m \frac{A_\ell}{2\ell+1}a_{\ell m} Y_{\ell m,n}\Delta\Omega+\sum_\ell \sum_m \frac{A_\ell}{2\ell+1}a_{\ell -m}(-1)^{2m} Y_{\ell -m,n}\Delta\Omega\nonumber\\
    &=&2\sum_\ell \sum_m \frac{A_\ell}{2\ell+1} a_{\ell m} Y_{\ell m,n}\Delta\Omega\,,
\end{eqnarray}
where $Y_{\ell m,n}$ is the spherical harmonic function for the $n$-th pixel and $\Delta \Omega$ is the solid angle of each pixel. Here we define $A_\ell \equiv (\rho_1 H_\ell+\lambda_\ell)/C_\ell^{\rm prior}$.} 


\section{Can {\tt AMIGO} still work beyond the Nyquist Limit?}
\label{app:aliasing}
\begin{figure*}[h!]
    \centering
    \includegraphics[width=0.32\columnwidth]{recon_noisefree_freq_3.0_nside_64.pdf}
    \includegraphics[width=0.32\columnwidth]{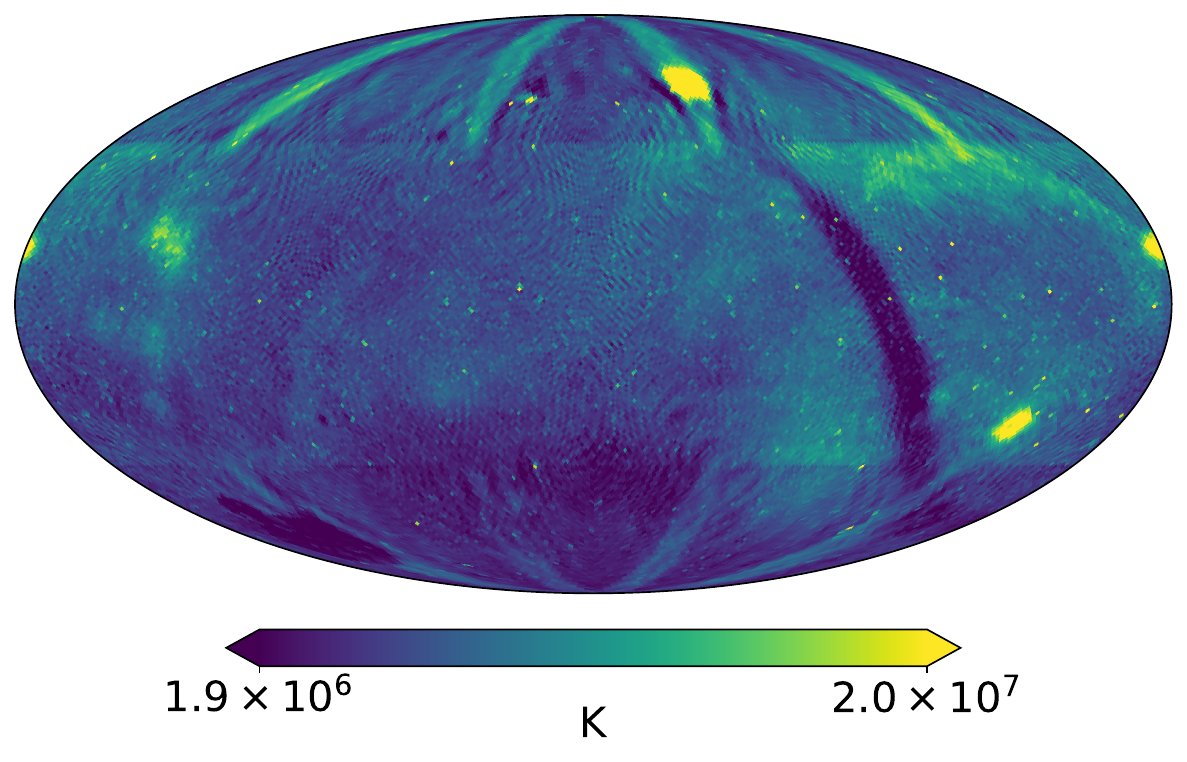}
    \includegraphics[width=0.32\columnwidth]{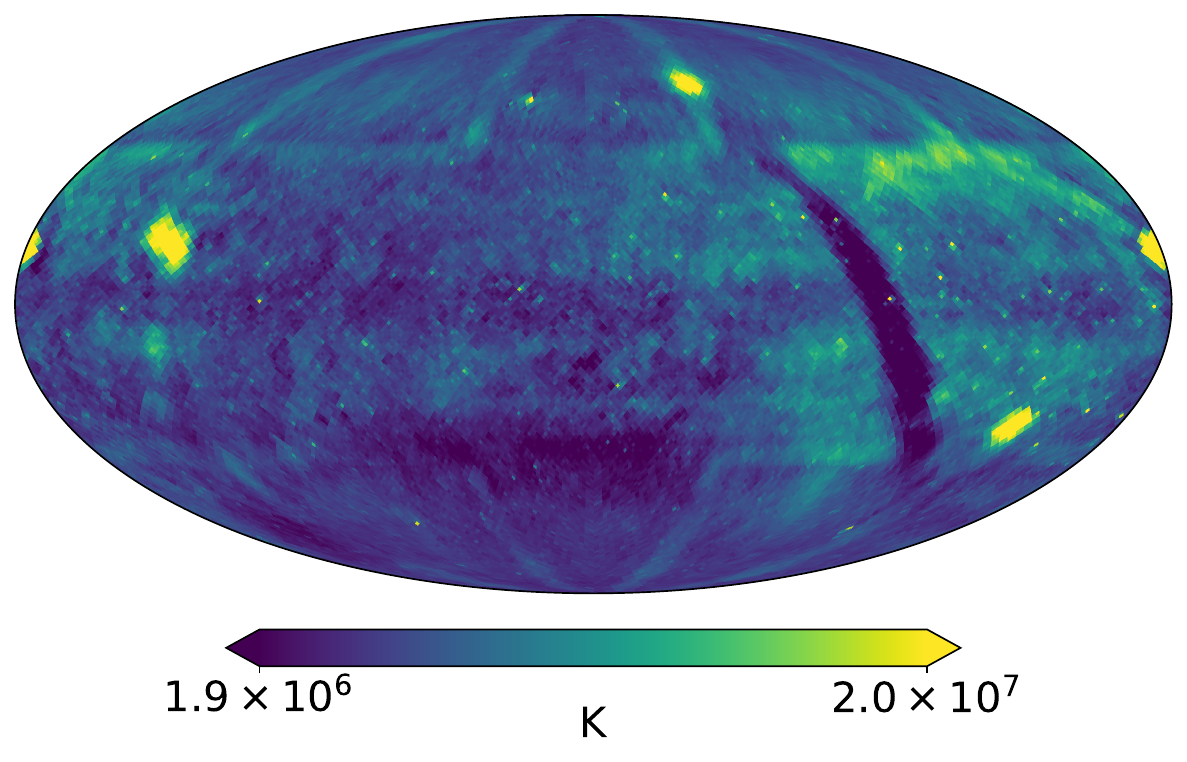}
    \caption{\revise{The reconstructed maps at 3.0 MHz with NSIDE = 64, using visibility data of different baseline ranges. We use visibility data of baselines $b<b_{\rm NQ}$ in the left panel, $b<2b_{\rm NQ}$ in the middle, and $b<4b_{\rm NQ}$ in the right. We find strong aliasing effects if the skymaps are reconstructed using visibility data beyond the Nyquist limit.}} 
    \label{fig:baselinerange}
\end{figure*}

\revise{For {\texttt{AMIGO}}, we restrict the used baselines to $b < b_{\rm NQ}$. Longer baselines sensitive to sub-pixel noise can introduce aliasing effects, as shown in Figure~\ref{fig:baselinerange}, which compares the reconstructed maps at 3.0 MHz using baselines with $b<b_{\rm NQ}$, $b<2\,b_{\rm NQ}$, and $b<4\,b_{\rm NQ}$ respectively. The values of their MSE and SSIM also decrease significantly, as given in Table~\ref{tab:cases}, verifying the failure of {\tt AMIGO} beyond the Nyquist limit.} 

\bibliography{opt}{}
\bibliographystyle{aasjournalv7}



\end{document}